\documentclass[utf8]{Frontiers_LaTeX_Templates/FrontiersinHarvard} 

\usepackage{url,hyperref,lineno,microtype,subcaption}
\usepackage[onehalfspacing]{setspace}

\usepackage{multirow}
\usepackage[normalem]{ulem}
\useunder{\uline}{\ul}{}
\usepackage[table,xcdraw]{xcolor}
\usepackage{stfloats}
\usepackage{array}    
\usepackage{makecell} 
\usepackage{booktabs} 

\def\keyFont{\fontsize{8}{11}\helveticabold }
\def\firstAuthorLast{Heldreth {et~al.}} 
\def\Authors{Courtney Heldreth\,$^{1,*}$, Diana Akrong\,$^{2}$, Laura M. Vardoulakis\,$^{3}$, Nicole E. Miller\,$^{4}$, Yael Haramaty\,$^{5}$, Lidan Hackmon\,$^{5}$, Lior Belinsky\,$^{5}$, Abraham Ortiz Tapia\,$^{4}$, Lucy Tootill\,$^{4}$, and Scott Siebert\,$^{4}$}
\def\Address{
$^{1}$Google Research, Seattle, WA, USA \\
$^{2}$Google Research, Accra, Ghana \\
$^{3}$Google Research, Mountain View, CA, USA \\
$^{4}$Bold Insight, Chicago, IL, USA \\
$^{5}$Google Research, Tel Aviv, Israel }
\def\corrAuthor{Courtney Heldreth}
\def\corrEmail{cheldreth@google.com}

\begin{document}
\onecolumn
\firstpage{1}

\title[An Experimental Evaluation of an AI-Powered Interactive Learning Platform]{An Experimental Evaluation of an AI-Powered Interactive Learning Platform} 

\author[\firstAuthorLast ]{\Authors} 
\address{} 
\correspondance{} 

\extraAuth{}

\maketitle

\begin{abstract}


Generative AI, which is capable of transforming static content into dynamic learning experiences, holds the potential to revolutionize student engagement in educational contexts. However, questions still remain around whether or not these tools are effective at facilitating student learning. In this research, we test the effectiveness of an AI-powered platform incorporating multiple representations and assessment through \textit{Learn Your Way}, an experimental research platform that transforms textbook chapters into dynamic visual and audio representations. Through a between-subjects, mixed methods experiment with 60 US-based students, we demonstrate that students who used \textit{Learn Your Way} had a more positive learning experience and had better learning outcomes compared to students learning the same content through a digital textbook. These findings indicate that AI-driven tools, capable of providing choice among interactive representations of content, constitute an effective and promising method for enhancing student learning.

\tiny
 \keyFont{ \section{Keywords:} Artificial Intelligence in Education, Personalized Learning, Content Transformations} 
\end{abstract}


\section{Introduction}
Around the world, AI is being leveraged to transform lives and augment human capabilities in multiple industries, and education is no exception. Teachers, students, and other stakeholders in the industry have been known to leverage AI to support lesson planning and delivery, learning, administrative tasks, etc. \citep{LUCZAK2024}. 

With the continuous push towards digital equity and literacy over the years, there has been an increased interest in education technologies (particularly chat-based tools like ChatGPT\footnote{https://chatgpt.com/} and Gemini\footnote{https://gemini.google.com/}) \citep{Zhai2021, ZHANG2021, Adeshola2024, MEMARIAN2023, Li2025}. Amongst other benefits, generative AI is uniquely poised to transform learning materials (including traditional textbooks) into more dynamic, engaging, and personalized content and to provide immediate feedback to improve understanding of learning materials \citep{LUCZAK2024}.

As improvements are made in natural language processing and understanding, large language models (LLMs) show promise for generating effective learning resources at scale for diverse groups of learners. This goes beyond performance outcomes to other intrinsic factors such as interest and motivation in learning. For example, in a between-subjects study comparing pre-existing text to AI-generated output, Leong et al.  show that while there was no difference in learning performance, AI-generated output positively affected learning motivation thereby facilitating positive downstream effects (including appreciating the relevance and utility of learning the material) \citep{Leong2024}.

Research in education also shows that learners benefit from having information presented in multiple modalities or formats \citep{mayer2002}. By presenting information in multimodal learning environments paired with guided activities, reflection, and feedback, learners benefit from numerous pedagogical advantages, including the promotion of independent learning and practice that is tailored to the needs of individual students \citep{Moreno2007}.

The process of transforming traditional textbooks and learning materials into these more dynamic, engaging, and personalized sources can, however, be time consuming and labor intensive for teachers, thereby limiting the broader use of AI-powered education tools \citep{dennison2025}. When considering direct and independent use by students for more personalized experiences at scale, several areas of concern emerge that demand careful consideration. These include the critical issues of accuracy, bias, academic integrity, factuality and reliability of model outputs, the potential for over-reliance on AI tools and its long-term impact on cognitive abilities and critical thinking skills, and students’ inherent capacity to formulate effective prompts for facilitating their learning experience \citep{han2024, Park2024, Tankelevitch2025, Chan2023, denny2023}. Furthermore, research has been mixed when it comes to demonstrating the efficacy of AI tools on performance outcomes \citep{do2025, Leong2024, Pallant2025}. 

In this paper, we build upon pedagogical best practices to experimentally test the impact of an AI-powered interactive learning platform on student understanding and recall of information. We investigate the following research questions:

\begin{itemize}
\item \textbf{RQ1} How does \textit{Learn Your Way}, an experimental research platform that leverages generative AI to transform content \citep{LYWtechreport}, impact student comprehension and immediate recall of information?
\item \textbf{RQ2} How does \textit{Learn Your Way} impact retention and recall of information over time?
\item \textbf{RQ3} Can \textit{Learn Your Way} facilitate more engaging learning experiences for students?
\end{itemize} 

To answer these questions, we conducted an experiment with 60 students aged 15-18 in the US, who were asked to learn a textbook chapter using a PDF reader or \textit{Learn Your Way}. We then examined learning outcomes, in addition to quantitative metrics of user sentiment and qualitative feedback about their learning experience.

The results of the experiment suggest that AI-powered interactive learning platforms that incorporate content transformations and assessments, can facilitate positive learning outcomes and provide students with more engaging learning experiences  compared to learning via traditional textbook chapters. We also demonstrate the positive role of timely feedback (as opposed to solutions) in equipping learners with knowledge and skills for self efficacy and academic independence.

\section{Background and Related Work}

\subsection{HCI and Education}
The Human-Computer Interaction (HCI) community has a rich history of research exploring the design, development, and evaluation of educational technology \citep{stasko1993, russell1993, abowd1998, scaife1997, zuckerman2005, Saerbeck2010, winkler2020}. The discourse about generative AI in education has shifted from an initial period of denial and uncertainty to partial acceptance and explorations, ultimately leading to calls for further research \citep{Pallant2025} and governance \citep{unicef2019ai}. Recent HCI research has explored a wide range of important topics in this domain, including privacy and security considerations for technology use in schools \citep{kumar2019,  Chanenson2023}, extending participatory and co-design methods to the development of classroom-based technologies \citep{chang2024, nicholson2022}, and designing tactile-based learning experiences to foster accessibility in education \citep{Melfi2020}. 

Additionally, research in HCI is beginning to explore how the increase in AI-powered capabilities can be incorporated into a variety of interactive experiences in order to positively impact learning. For example, Liu et al. created and evaluated an AI agent designed to interact with students in VR-based learning environments and promote classroom participation and more positive learning experiences, finding that these agents can provide pedagogical value through increased levels of engagement and the potential for learning gains \citep{Liu2024}. Chen et al. found that augmenting block-based programing software (Scratch) with an LLM-based chat experience allowed students to learn more independently, and create higher quality projects \citep{Chen2024}.  This growing body of research suggests a strong desire from the HCI community to better understand how generative AI tools can help improve student outcomes. 

\subsection{AI Assessment and Feedback}
One of the ways AI can help bring value to students is through AI-driven assessment and feedback. Pedagogical research shows that effective feedback is timely, actionable, grounded in standards or learning goals, and cites specific observations to facilitate learner understanding and support action \citep{wiggins2012seven}. Providing feedback is an inherent part of a teacher’s role, but as students partake in independent practice outside of classroom hours, feedback is not always readily available or accessible. However, the ability to access feedback is necessary to help students understand their performance and progress in relation to required standards and learning objectives \citep{hattie2007}.  

From the early days of intelligent tutoring systems, AI-driven knowledge assessments – including personalized, adaptive feedback – has been a core area of research for education-focused AI \citep{anderson1985, corbett1997, graesser2004autotutor}. With generative AI, students have the potential to access ongoing, targeted, timely, and individualized assessments and feedback at scale for self-directed learning and self-efficacy. Recent technological advances show improvements with the capabilities surrounding assessment and feedback \citep{Choi2020, yang2022}, but more improvements are needed to truly reach this vision \citep{tu2025empowering}. 

In HCI, several qualitative studies with parents, teachers, and students have led to  design recommendations that emphasize the value, importance, and opportunity for AI to provide students with high-quality assessments and feedback \citep{han2024, prasad2025}. A recent meta-analysis on emerging technologies in education also emphasizes the importance of HCI research that is grounded in pedagogical strategies and focuses on assessment and feedback \citep{vanmechelen2023}. However, there is limited quantitative data and experimental studies evaluating how the \textit{design} of AI-powered education tools can facilitate effective assessments and feedback. In this paper, we build upon prior work by incorporating design recommendations from past research into an experimental generative AI platform, \textit{Learn Your Way}, and subsequently assessing the platform's impact on student learning experiences and performance outcomes. 

\subsection{AI-Generated Content Transformations}
By definition, content transformations transform static content into multiple representations to help facilitate learning, such as transforming text-based content into something that is audio- and/or visual-based \citep{AINSWORTH1999}. Content transformations are an integrated process that isn't just changing the modality (i.e., text to audio), but also the structural reorganization of information that bridges the gap between raw information and a learner's cognitive architecture \citep {shulman1987knowledge, Koedinger2015}. In contemporary educational technology and HCI research, there is a notable rise in the adoption of generative content transformation capabilities; these extend beyond simple modality shifts to include the structural transmutation of static information into interactive assessment tools and adaptive feedback mechanisms \citep{brusilovsky2001adaptive, mitrovic2012fifteen}. This trend reflects a broader shift towards dynamic, personalized learning resources that can be adapted to various student needs and learning preferences. The benefits of content transformations are guided by principles from Mayer’s cognitive theory of multimedia learning, which states that people learn more deeply when information is received in multiple modalities as opposed to reading or hearing words alone \citep{mayer2002,  mayer1998cognitive}. For example, a study showed that children who listened to stories while also viewing relevant pictures recalled more information than those who only heard stories \citep{levin1980children}. This study (and others, e.g., \citep{Villegas2024}) suggests that integrating materials across different modalities, such as audio paired with images or videos, can be particularly beneficial for student engagement and learning. 

Generative AI capabilities, by their nature, create content transformations (text-to-text, text-to-audio, text-to-image, and text-to-video). By providing multiple modes of interaction and engagement with content, generative AI tools are well positioned to support self-directed learning, including theories that advocate for learners to play a central role in their own learning journeys, which has been shown to have positive impacts on learning outcomes  \citep{Deci2000, Zimmerman2009, zimmerman2000}. For example, this could include metacognitive strategies that allow for setting individual learning goals, choosing appropriate strategies to achieve these goals, and monitoring one’s individual learning process and outcomes for adjustment and improvements \citep{Panadero2017}. 

In HCI, the research community is beginning to explore the effects of AI-based content transformations on learning and education. In an experiment, Leong et al. found that a generative AI-based vocabulary app, that transformed standard vocabulary sentences into either personalized, generated sentences or personalized, generated text-based stories did not increase learning performance but did positively impact learning motivation \citep{Leong2024}. Similarly, in an experiment moving beyond text-to-text transformations, Do et al. found that transforming educational materials from text to podcast form led to a more enjoyable learning experience and also, significantly improved learning outcomes for certain subjects \citep{do2025}. 

Taken together, the latest research in HCI suggests that AI-based content transformations are poised to enhance learning experiences. However, more research is needed to understand the potential of AI in education by studying a broader range of the content transformation capabilities available through generative AI. Additionally, while a majority of prior HCI studies have explored model-driven interaction paradigms, there is also a notable gap in research that allows for user-driven experiences. This study addresses this gap through \textit{Learn Your Way}, a system that extends beyond singular text and audio transformations, and provides learners with five different learning modalities that leverage multiple content transformations to create an interactive, engaging experience that enables user agency and self-direction, as well as real-time assessment and feedback.  

\section{Learn Your Way}

\textit{Learn Your Way} is an experimental platform that leverages generative AI to transform textbook PDFs into a variety of formats, or multiple representations \citep{LYWtechreport}. We used \textit{Learn Your Way} to test our research hypotheses. \textit{Learn Your Way} is not intended to be an educational product used in classrooms. Rather, it is a way for us to test how AI and pedagogical theories have the potential of impacting learning outcomes. These content transformations were grounded in pedagogical theories that demonstrate that multiple representations of content can be a powerful way to learn \citep{Mayer1990, mayer1998cognitive}. \textit{Learn Your Way also} incorporates learner control over format selection, a feature theoretically grounded in Self-Determination Theory (SDT) \citep{ryan2000self}. By enabling this choice, we aim to augment the students' perceived autonomy, a known precursor to enhancing intrinsic motivation, cognitive effort allocation, task performance, and self-efficacy (or perceived competence) \citep{Reber2009, SCHNEIDER2018, patall2008effects}.

\textit{Learn Your Way} presents five learning modalities: Immersive Text, Slides, Video, Audio Lesson, and Mindmap (Figure \ref{fig:figure1}). See Supplementary Materials for a visual example of each modality. 

\begin{itemize}
    \item \textbf{Immersive Text}:  Transforms the original content into smaller sections releveled to a selected grade, and provides opportunities for interactive learning and practice (e.g. memory aids, quizzes, embedded questions, and timelines). These transformations of chunking and scaffolding are not just “extra features,” but mechanisms by which AI generates  “dynamic” content rather than “static.” Within Immersive Text, students had access to the following features:
     \begin{itemize}
        \item \textbf{Section-level quizzes (QuizMe)}: Interactively quizzes the learner to reinforce and assess what was just learned and uncover existing knowledge gaps 
        \item \textbf{Embedded questions}: Short, in-line questions to check for understanding while the learner is reading
        \item \textbf{AI-Generated Images and Summary (Enimate)}: Briefly explains key concepts through visuals and short descriptions 
        \item \textbf{Timeline}: AI-generated visualization of the sequence of key concepts in the original content
        \item \textbf{Memory Aids}: AI-generated mnemonic that assists in memorization and recall of key concepts in the original content
   \end{itemize}
   \item \textbf{Slides}: Comprehensive presentations that cover the entire source material and include reflection prompts like fill-in-the-blanks. 
   \item \textbf{Video (slide deck with narration)}:  Provides the option for a fully narrated slideshow mimicking a recorded lesson.
   \item \textbf{Audio Lesson}: Simulated conversations between an AI-powered teacher and student that models how a real learner might engage with the material. This virtual student asks questions and can even express common misconceptions which the teacher clarifies. The conversation is accompanied by visual diagrams to further illustrate the concepts.
   \item \textbf{Mindmap}: Organizes the knowledge hierarchically into an interactive diagram that allows the learner to zoom in and out from the big picture to the minute details.
 \end{itemize}

\begin{figure}[b]
\centering
\includegraphics[width=0.99\columnwidth]{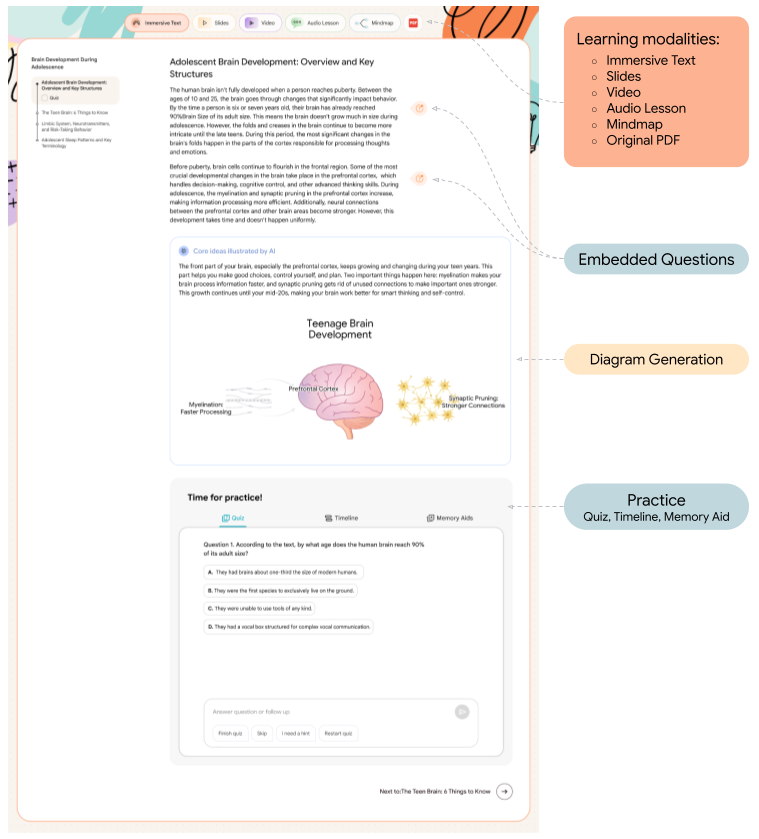}
  \caption{Diagram of the \textit{Learn Your Way} user interface for Immersive Text. Other learning modalities are featured at the top including Slides, Video, Audio Lesson, and Mindmap, as well as the original PDF.}
  \label{fig:figure1}
\end{figure}

When students open \textit{Learn Your Way}, they start by selecting their grade level and interests (see Appendix \ref{appA}). \textit{Learn Your Way} then personalizes and adapts the content being generated by adjusting for grade level and incorporating the student’s selected interests. As students use \textit{Learn Your Way}, they engage with interactive quizzes designed to promote active learning and assess real-time understanding of the source material, enabling further elements of personalization. Students receive dynamic feedback on their answers, can track their overall progress, and are guided to review specific content areas where they demonstrated gaps in understanding. Students are also able to view the original content, in PDF form, at any time. 

\subsection{Learn Your Way Design and Evaluation}
Throughout the design of \textit{Learn Your Way}, each modality was evaluated by pedagogical experts to ensure the fidelity and quality of each content transformation while preserving the integrity of the source content. A unified rubric was developed to assess key pedagogical criteria, including accuracy (the extent to which the AI faithfully represented the source material), coverage (the thoroughness with which all key concepts were addressed), and learning science principles.
The results indicated a high level of quality, with all modalities and features achieving an average expert rating of 0.9 or higher on a scale of 0 to 1 across all pedagogical criteria. Detailed information about the rubric and results of the evaluation can be found in Supplementary Materials.

\section{Experimental Study}
We conducted an experimental study to compare performance outcomes and learning experiences between the AI-powered learning platform (\textit{Learn Your Way}) and a digital reading of an original textbook chapter (\textit{Textbook}) as a baseline comparison, consistent with established baseline comparisons employed in the literature \citep{do2025}. 

The source content used in the experiment was a neuroscience textbook chapter titled, “Brain Development During Adolescence” \citep{lumen_brain_adolescence}. The source material was chosen by a panel of pedagogy experts using a set of criteria based on education psychology (see Supplementary Materials). 
\subsection{Research Hypotheses}
We present the following research hypotheses: 
\begin{itemize}
    \item [\textbf{H1}] We hypothesize that participants will perform better on both an immediate and long-term recall assessment in the \textit{Learn Your Way} condition compared to the \textit{Textbook} condition, drawing on prior research demonstrating that transforming static content into different modalities and self-directed learning can  have a positive effect on learning outcomes \citep{AINSWORTH1999, do2025, Deci2000}.
    \item [\textbf{H2}] Consistent with research that demonstrates that students experience enjoyment from AI-generated tools \citep{Abdaljaleel2024, Do2022, do2025, Leong2024}, we hypothesize that participants in the \textit{Learn Your Way} condition will report more positive learning experiences compared to those in the \textit{Textbook} condition. 
\end{itemize}

\subsection{Dependent Variables}
All assessments and questionnaires were programmed and administered through Qualtrics\footnote{www.qualtrics.com}. Full question banks can be found in the Supplementary Materials.

\subsubsection{Learning Outcomes}
To assess immediate recall and comprehension, participants completed a  $\sim$10 minute, in-session assessment that consisted of short answer questions (SAQs), single-answer multiple choice questions (SMCQs), multi-answer multiple choice questions (MMCQs), and matching questions. The immediate recall assessment was scored out of 12 points.

To assess long-term recall, participants were sent a brief follow-up assessment three days post-session that consisted of one short answer question, one single-answer multiple choice question, and one matching question designed to reassess the learning outcomes targeted by the initial, immediate assessment. The long-term recall assessment was scored out of 6 points.

Both assessments were developed by pedagogical experts and targeted specific levels of Bloom’s Taxonomy in order to ensure cognitive complexity  \citep{bloom1971taxonomy}. The assessments were therefore designed to be a good measure of content comprehension and long-term knowledge consolidation (see Supplementary Materials for all assessment questions). 

In order to establish consistency of SAQ scoring across both assessments, one of our pedagogy experts pre-defined rubrics with scoring criteria and examples for each SAQ on a 3-point scale. All SAQ responses were evaluated solely on students' demonstrated knowledge of the source material (ranked from \textit{0-Fragmentary} to \textit{3-Demonstrating}), excluding any consideration of students' grammar skills or writing proficiency in this evaluation (see Supplementary Materials for complete rubrics). Our pedagogy expert and a member of the research team independently scored and compared results for one third of the data to ensure alignment and adherence to the rubric guidelines. Both were blind to the experimental condition. The remainder of SAQ scoring was completed independently by the study’s pedagogy expert.

\subsubsection{Learning Experience}
Following the immediate recall assessment, participants responded to a series of survey questions that assessed their experiences recalling the content and using the educational tool. These questions included ratings of participants’ familiarity and interest in the topic, perceived ease / difficulty of the assessment, and the utility the tool provided in helping the participant learn the content. We also asked a series of agreement statements that evaluated the tool’s effectiveness, desire to use in the future, and enjoyment.

\begin{figure*}[h!t]
\centering
\includegraphics[width=0.95\textwidth]{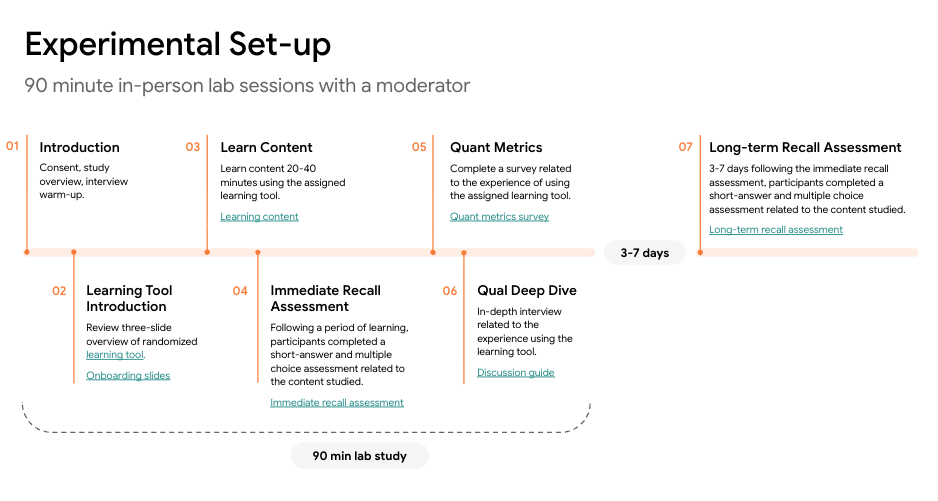}
    \caption{Overview of the experimental set-up.}
    \label{fig:ExpSetupFig2}
\end{figure*}

\subsection{Recruitment}
Participants were recruited through a research vendor (Bold Insight\footnote{https://boldinsight.com/}). During recruitment, participants completed a screener survey that collected information on their age, grade level, GPA, standardized test scores, and relevant academic courses taken.  

In order to evaluate potential biasing effects of prior familiarity with the learning material on participant feedback and assessment scores, we had participants rate their familiarity across 15 age-appropriate topics, including the topic of the chapter featured in the experiment – adolescent brain development. Across participants, baseline familiarity with the selected learning topic was highly clustered around the mid-point of the scale ($Mdn=3$, $M=2.7$, $SD=1.2$) suggesting partial but not high familiarity with the learning material. Furthermore, after random assignment, baseline content familiarity was approximately matched across experimental conditions.

To ensure students participating in the study had similar levels of reading comprehension, they read a short passage and completed a reading comprehension task during the recruitment process. The average score on the assessment was 6.4 out of 10, with a standard deviation of 2.3. We included students who scored one standard deviation above or below the mean, or those who scored a 4-9 out of 10 on the reading assessment.

\subsection{Procedure}
The experiment consisted of a single 90 minute in-person lab session and a follow-up assessment shared remotely with participants three days post-session. See Figure \ref{fig:ExpSetupFig2} for an overview of the experimental setup.  The study was conducted in adherence to Google and Bold Insight's ethical, legal, and privacy standards for human subjects research, and our participants completed an informed consent process.

At the beginning of the study, participants were randomly assigned to one of the two experimental conditions (\textit{Learn Your Way} or \textit{Textbook}) and were given a brief overview (three slides) that introduced the educational tool they would be using during the task (see Supplementary Materials). They were then asked to imagine they were sitting in one of their classrooms at school where their teacher asked them to learn the educational content using the provided tool and complete an assessment based on what they learned. Participants in the \textit{Learn Your Way} condition selected a preloaded PDF of the learning material and indicated the grade level they wanted to learn the content in, as well as their interests (e.g., music, video games, biking, etc). All participants in the \textit{Learn Your Way} condition started in the Immersive Text modality, and therefore interacted with the QuizMe feature. Participants in the \textit{Textbook}  condition viewed the chapter in a web-based PDF reader. To ensure uniformity across conditions, all participants were given up to 40 minutes (but required to spend at least 20 minutes) using the tool and learning the content before taking a short assessment. We report the maximum and average learning duration by educational tool condition in Table \ref{tab:table1}.

\begin{table*}[t]
\centering
\caption{Duration of participant tool usage for \textit{Learn Your Way} and \textit{Textbook}.}
\vspace*{2mm} 
\label{tab:table1}
\renewcommand{\arraystretch}{1.5} %
\begin{tabular}{p{8cm}cc}
\toprule
&
\multicolumn{2}{c}{\textbf{Tool Condition}} \\ \cline{2-3} 
\textbf{Duration of tool use} & 
\textit{\textbf{Learn Your Way}} & 
\textit{\textbf{Textbook}} \\ \hline
Maximum time used (minutes and seconds) & 40:50  & 32:49 \\
Average time used (minutes and seconds) & 30:49  & 22:05     \\

\end{tabular}
\end{table*}

After the allotted 40 minutes (or when the participant indicated they were ready to move on), participants completed the immediate assessment ($\sim$10 minutes) and the learning experience questionnaire ($\sim$5 minutes) via a digital survey. Finally, they spent 30 minutes in a qualitative, debrief interview with a skilled moderator. In these in-depth interviews, participants were asked questions about their experience using the educational tool, utility of the tool and corresponding capabilities, experience with the assessment, recommendations for improving the tool, and how the tools compare to current tools they use for learning. 

Three days post-session, participants were emailed a follow-up assessment that measured long-term recall of the content they had learned during the experiment. 58 out of 60 participants completed the long-term recall assessment within the required 3--7 day window. Participants were compensated between \$115--\$160, accounting for study time, travel time, and completion of the follow-up survey.

\begin{table*}[b]
\centering
\caption{Participant demographics by educational tool condition, location, and grade level.}
\vspace*{2mm} 
\label{tab:table2}
\renewcommand{\arraystretch}{1.5} %
\begin{tabular}{m{5cm}m{3cm}rrrrr}
\toprule
&
&
\multicolumn{5}{c}{{\color[HTML]{1B1C1D} \textbf{\begin{tabular}[c]{@{}c@{}}Number of Participants\\ Grade Level\end{tabular}}}} \\ \cline{3-7} 
\multirow{-2}{*}{\textbf{Tool Condition}} &
  \multirow{-2}{*}{\textbf{Location}} &
  \multicolumn{1}{c}{{\color[HTML]{1B1C1D} \textbf{9th}}} &
  \multicolumn{1}{c}{{\color[HTML]{1B1C1D} \textbf{10th}}} &
  \multicolumn{1}{c}{{\color[HTML]{1B1C1D} \textbf{11th}}} &
  \multicolumn{1}{c}{{\color[HTML]{1B1C1D} \textbf{12th}}} &
  \multicolumn{1}{l}{{\color[HTML]{1B1C1D} \textbf{Grand Total}}} \\ \hline
{\color[HTML]{1B1C1D} } &
  {\color[HTML]{1B1C1D} Urban} &
  {\color[HTML]{1B1C1D} 1} &
  {\color[HTML]{1B1C1D} 3} &
  {\color[HTML]{1B1C1D} 4} &
  {\color[HTML]{1B1C1D} 2} &
  {\color[HTML]{1B1C1D} 10} \\
{\color[HTML]{1B1C1D} } &
  {\color[HTML]{1B1C1D} Suburban} &
  {\color[HTML]{1B1C1D} 0} &
  {\color[HTML]{1B1C1D} 2} &
  {\color[HTML]{1B1C1D} 7} &
  {\color[HTML]{1B1C1D} 2} &
  {\color[HTML]{1B1C1D} 11} \\
\multirow{-3}{*}{{\color[HTML]{1B1C1D} \textit{\textbf{Learn Your Way}}}} &
  {\color[HTML]{1B1C1D} Rural} &
  {\color[HTML]{1B1C1D} 0} &
  {\color[HTML]{1B1C1D} 3} &
  {\color[HTML]{1B1C1D} 5} &
  {\color[HTML]{1B1C1D} 1} &
  {\color[HTML]{1B1C1D} 9} \\
\rowcolor[HTML]{EFEFEF} 
{\color[HTML]{1B1C1D} \textbf{Total -} \textit{\textbf{Learn Your Way}}} &
   &
  {\color[HTML]{1B1C1D} 1} &
  {\color[HTML]{1B1C1D} 8} &
  {\color[HTML]{1B1C1D} 16} &
  {\color[HTML]{1B1C1D} 5} &
  {\color[HTML]{1B1C1D} 30} \\ \hline
{\color[HTML]{1B1C1D} } &
  {\color[HTML]{1B1C1D} Urban} &
  {\color[HTML]{1B1C1D} 1} &
  {\color[HTML]{1B1C1D} 1} &
  {\color[HTML]{1B1C1D} 4} &
  {\color[HTML]{1B1C1D} 2} &
  {\color[HTML]{1B1C1D} 8} \\
{\color[HTML]{1B1C1D} } &
  {\color[HTML]{1B1C1D} Suburban} &
  {\color[HTML]{1B1C1D} 2} &
  {\color[HTML]{1B1C1D} 5} &
  {\color[HTML]{1B1C1D} 7} &
  {\color[HTML]{1B1C1D} 0} &
  {\color[HTML]{1B1C1D} 14} \\
\multirow{-3}{*}{{\color[HTML]{1B1C1D} \textit{\textbf{Textbook}}}} &
  {\color[HTML]{1B1C1D} Rural} &
  {\color[HTML]{1B1C1D} 0} &
  {\color[HTML]{1B1C1D} 4} &
  {\color[HTML]{1B1C1D} 3} &
  {\color[HTML]{1B1C1D} 1} &
  {\color[HTML]{1B1C1D} 8} \\
\rowcolor[HTML]{EFEFEF} 
{\color[HTML]{1B1C1D} \textbf{Total -} \textit{\textbf{Textbook}}} &
   &
  {\color[HTML]{1B1C1D} 3} &
  {\color[HTML]{1B1C1D} 10} &
  {\color[HTML]{1B1C1D} 14} &
  {\color[HTML]{1B1C1D} 3} &
  {\color[HTML]{1B1C1D} 30} \\ \hline
\rowcolor[HTML]{EFEFEF} 
{\color[HTML]{1B1C1D} \textbf{Grand Total}} &
   &
  {\color[HTML]{1B1C1D} 4} &
  {\color[HTML]{1B1C1D} 18} &
  {\color[HTML]{1B1C1D} 30} &
  {\color[HTML]{1B1C1D} 8} &
  {\color[HTML]{1B1C1D} 60} \\
\end{tabular}
\end{table*}

\subsection{Participants}
We recruited 60 high school students from the Chicago area in the United States. We 
intentionally recruited students from rural ($n=17$), urban ($n=18$), and suburban ($n=25$) areas to ensure diversity in our participant pool. 52\% of participants were male, 47\% female, and 1\% non-binary. All participants were pre-screened to ensure they had similar levels of reading comprehension and did not require additional test-taking accommodations. 

Participants were randomly assigned to one of two educational tool conditions: \textit{Learn Your Way} (treatment condition) or \textit{Textbook} (control condition). Table \ref{tab:table2} shows a breakdown of participants' demographics across conditions.

\subsection{Data Analysis and Approach}
To address our research questions, we analyzed the collected data at multiple levels.  All data and statistical analyses were conducted in R and RStudio including the statistical package \textit{coin} for our non-parametric tests \citep{Hothorn2008} and package \textit{mgcv} for the generalized additive models \citep{wood2017gam}.

First, to inform the statistical method for immediate recall and long-term recall, we conducted a Shapiro-Wilk test on each dependent variable to check for the parametric assumption of a normal distribution \citep{Shapiro1965}. We addressed any violations of this assumption by conducting a non-parametric alternative to the independent t-test, the Wilcoxon-Mann-Whitney test. Specifically, we conducted one-sided Wilcoxon-Mann-Whitney tests on both immediate recall assessment scores and long-term recall assessment scores, in line with our first research hypothesis (H1). To address our second hypothesis (H2), stating that \textit{Learn Your Way} participants will report more positive learning experiences than \textit{Textbook} participants, we conducted directional Wilcoxon-Mann-Whitney tests for each of our learning experience metrics. Additionally, to control for the inflation of Type I error rate due to multiple comparisons, $p$-values were adjusted for the learning experience metrics using the Benjamini-Hochberg procedure to maintain a false discovery rate (FDR) of .05 \citep{benjamini1995controlling}. For all statistical tests, we report the \textit{Z}-statistic, \textit{p}-value, adjusted \textit{p}-value (where applicable), and subsequent effect size \textit{r}. 

Second, as an exploratory analysis, we investigated immediate recall and long-term recall performance for \textit{Learn Your Way} participants as a function of learning modality usage. We report mean values and directional usage patterns for three levels of modality usage: Immersive Text only, Immersive Text plus one additional modality, and Immersive Text plus 2 or more learning modalities utilized by \textit{Learn Your Way} participants. To increase sample and our subsequent power to detect differences, we combined the latter two usage types for our statistical analyses including two-sided Wilcoxon-Mann-Whitney tests and Pearson's correlation tests on both immediate recall and long-term recall scores.

In addition, we conducted several bias checks to assess whether extraneous factors may be driving the primary effects we observed. To explore whether students’ prior reading comprehension levels influenced our learning outcomes, we compared screener assessment scores for \textit{Learn Your Way} participants to those for \textit{Textbook} participants through a two-sided Wilcoxon-Mann-Whitney test on a reading comprehension task (eligible scores for participation ranging from 4--9 points). Next, to ensure the learning material struck a balance of being relatively unfamiliar to participants (as to not bias their recall scores), but still interesting enough to encourage participant engagement in the task, we had them rate their level of interest and familiarity with the content. Participants' baseline familiarity was evaluated in two ways: first, through a screener questionnaire and again, during the in-session survey. Both content familiarity and interest ratings were analyzed using two-sided Wilcoxon-Mann-Whitney tests, appropriately suited for ordinal data. For our final bias check, we conducted a two-sided Wilcoxon-Mann-Whitney test on students' perceived assessment difficulty, rated on a standard five-point scale from ``very difficult" to ``very easy." 

To further ensure that the observed learning outcomes were not driven by procedural confounds, we modeled the influence of learning duration (transformed into seconds) and delay duration (within the 3--7 day window) on assessment scores. Specifically, we employed semi-parametric generalized additive models (GAMs) to 1) allow for potential non-linear dependencies between learning duration and recall performance, and 2) realistically model the non-linear memory decay typically associated with the forgetting curve \citep{ebbinghaus1913memory}. Within these models, smoothing terms were applied to both continuous predictors -- learning duration and delay duration -- nested by experimental condition, to allow the models to optimally fit both linear and non-linear trends. The complexity of these smooth terms was evaluated using effective degrees of freedom ($edf$), where $edf = 1$ corresponds to a linear relationship and $edf > 1$ indicates increasing non-linearity \citep{wood2017gam}.    

Finally, for our qualitative interviews, we applied Braun and Clarke’s thematic analysis method \citep{braun2021thematic} to generate themes that helped us understand the key patterns observed in the quantitative data in more depth. We chose this approach because it is widely used in other HCI interview-based studies \citep{Johnson2022, Wong2025}. A data collection tool was created in Google Sheets to capture participant behavior and interview data.  During each session, the moderator used a paper guide to facilitate the interview and take notes while a notetaker entered participant responses in the data collection tool.  Post-session, the moderator and notetaker reviewed the study data for accuracy and completeness. As needed, session videos and transcripts were also reviewed to confirm study data accuracy. The study team debriefed daily to review, discuss, and refine ongoing themes from the interviews.  Each week, the team also met to discuss the study data and further align on trends and themes observed. The team met to refine the analysis until data saturation was achieved \citep{hennink2017code} resulting in a final set of themes focused on differences in students’ behavior, attitudes, and reflections on their learning experiences.

\section{Results}

\subsection{Bias Checks}
A Shapiro-Wilk test confirmed a significant departure from normality for reading comprehension scores ($N= 60, W= 0.93, p= .002$). We therefore conducted a non-parametric Wilcoxon-Mann-Whitney test and found no significant difference in scores between \textit{Learn Your Way} ($Mdn= 7; M= 6.5, SD= 1.4$) and \textit{Textbook} students ($Mdn= 7; M= 6.7, SD= 1.4$) on the reading comprehension task  ($Z=0.54$, $p=.588$, $r=.07$). 

In addition, we explored whether prior familiarity with the learning material differed by experimental condition for either the screener ratings or in-session ratings of familiarity. We found no significant difference in baseline familiarity with the learning material between \textit{Learn Your Way} ($Mdn=2.5$; $M=2.6$, $SD=1.1$) and \textit{Textbook} conditions ($Mdn= 3$; $M= 2.9$, $SD= 1.2$) for screener ratings of familiarity ($Z=0.98$, $p=.337$, $r=.13$). Additionally, no significant difference was detected between the \textit{Learn Your Way} condition ($Mdn=3$; $M=2.6$, $SD=0.9$) or the \textit{Textbook} condition ($Mdn=2.5$; $M=2.4$, $SD=0.9$) for content familiarity assessed during the in-session questionnaire ($Z=0.52$, $p=.616$, $r=.07$). We also found no significant difference between the \textit{Learn Your Way} ($Mdn=3$; $M=3.3$, $SD=0.6$) and \textit{Textbook} conditions ($Mdn=3$; $M=3.2$, $SD=0.8$) in terms of students' reported interest in the chapter topic ($Z=0.11$, $p=.934$, $r=.01$). In fact, the majority of participants, regardless of experimental condition, rated the learning content as ``somewhat interesting" or ``very interesting" on a 4-point Likert scale.

Lastly, we evaluated students’ perceived assessment difficulty, as it was important for the immediate recall assessment to feel like a reasonable task for high school students – in that it was not too difficult nor too easy. We found no significant difference in student ratings of assessment ease or difficulty between the \textit{Learn Your Way} condition ($Mdn=4$; $M=3.6$, $SD=0.6$) and the \textit{Textbook} condition ($Mdn=3$; $M=3.4$, $SD=0.8$) after conducting a two-sided Wilcoxon-Mann-Whitney test ($Z=1.15$, $p=.268$, $r=.15$). 

These bias checks confirm that students’ reading comprehension levels, familiarity and interest in the topic, and perceptions of assessment difficulty did not meaningfully differ between experimental conditions, thereby providing no differential benefit to learning outcomes or learning experiences.

\subsection{Learning Outcomes}
In support of our first hypothesis (H1), we found that students in the \textit{Learn Your Way} condition ($Mdn=9.5$; $M=9.2$, $SD=2.2$) performed significantly better on the immediate recall assessment compared to students in the \textit{Textbook} condition ($Mdn=8$; $M=8.2$, $SD=2$).

In addition, analysis of long-term recall scores provides further support for H1, with \textit{Learn Your Way} participants outperforming \textit{Textbook} participants on the 3-7 day post-session assessment. Students in the \textit{Learn Your Way} condition ($Mdn=5$; $M=4.7$, $SD=1.2$) scored significantly higher on the long-term recall assessment than students in the \textit{Textbook} condition ($Mdn=4$; $M=4$, $SD=1.4$). 

\begin{table}[h]
\centering
\caption{Median (\textit{Mdn}), Interquartile range (\textit{IQR}), and sample size (\textit{n}) of immediate recall and long-term recall assessment scores by educational tool condition. The immediate recall assessment was scored out of 12 points while the long-term recall assessment was scored out of 6 points.}
\vspace*{3mm} 
\label{tab:table3}
\renewcommand{\arraystretch}{1.5} %
\begin{tabular}{m{5cm}cc}
\toprule
\textbf{Tool Condition} &
  \begin{tabular}[c]{@{}c@{}}\textbf{Immediate Recall}\\ \textit{Mdn}, \textit{IQR} (\textit{n})\end{tabular} &
  \begin{tabular}[c]{@{}c@{}}\textbf{Long-term Recall}\\ \textit{Mdn}, \textit{IQR} (\textit{n})\end{tabular} \\ \hline
\textit{\textbf{Learn Your Way}} &
  9.5, 3 (30) &
  5.0, 2 (29) \\
\textit{\textbf{Textbook}} &
  8.0, 2 (30) &
  4.0, 2 (29) \\

\end{tabular}
\end{table}

\begin{figure}[t]
\centering
  \includegraphics[width=0.99\columnwidth]{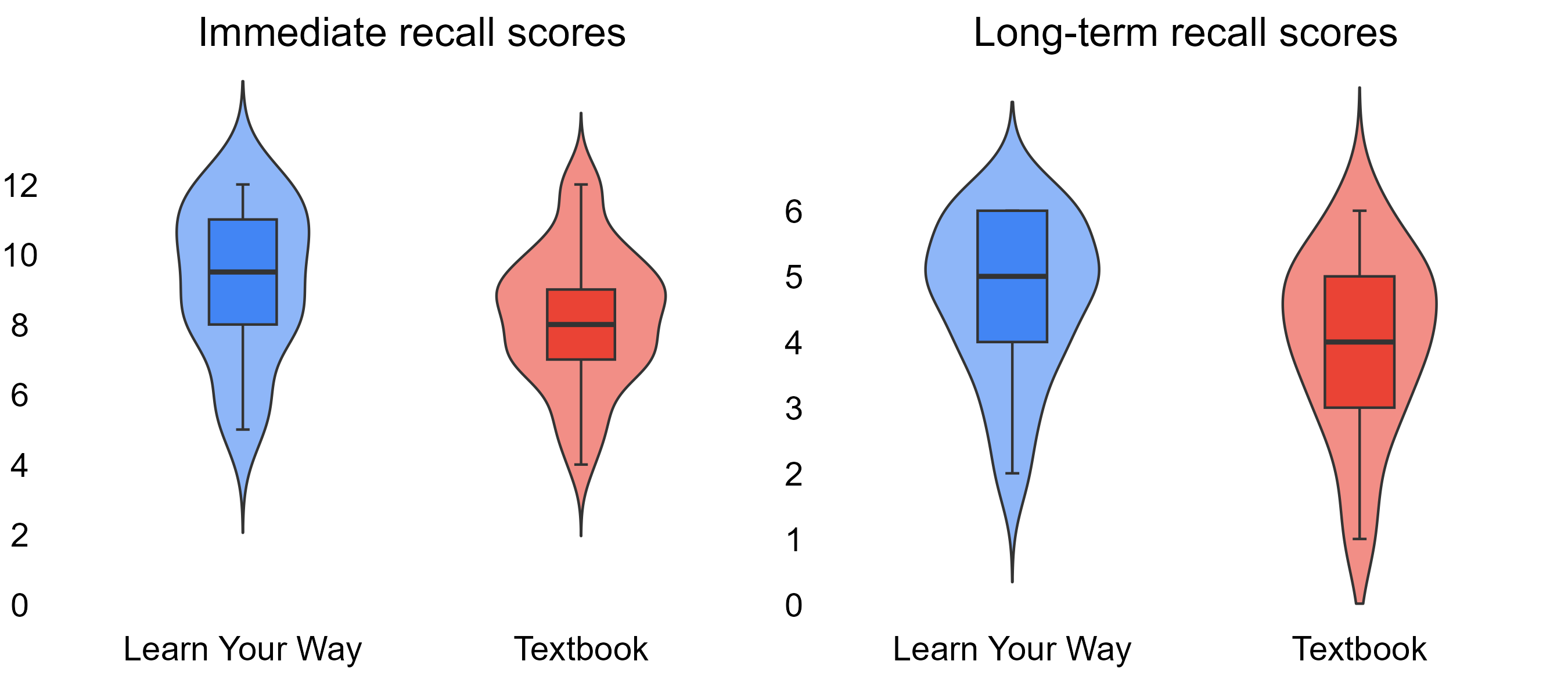}
  \caption{Violin and boxplots of immediate recall scores (left plot) and long-term recall scores (right plot) by tool condition. The immediate recall assessment was scored out of 12 points while the long-term recall assessment was scored out of 6 points.}
  \label{fig:figure3}
\end{figure}

In Table \ref{tab:table3}, we further breakdown immediate and long-term recall performance, including median scores and interquartile range (\textit{IQR}), and in Table \ref{tab:table4}, we show the results of statistical tests performed. Finally, we use violin and box plots to represent the overall score distributions for \textit{Learn Your Way} and \textit{Textbook} in Figure \ref{fig:figure3}.

\begin{table}[h]
\centering
\caption{Summary of Wilcoxon-Mann-Whitney tests of immediate and long-term recall scores with directional hypothesis testing. Asterisks denote significant differences at \textit{p}$<$.05\textsuperscript{*}. Effect sizes are reported as the rank-biserial correlation.}
\vspace*{3mm} 
\label{tab:table4}
\renewcommand{\arraystretch}{1.5} %
\begin{tabular}{m{5cm}cc}
\toprule
\textbf{Directional hypothesis} &
  \begin{tabular}[c]{@{}c@{}}\textbf{Immediate Recall} \\ Test statistics\end{tabular} &
  \begin{tabular}[c]{@{}c@{}}\textbf{Long-term Recall} \\ Test statistics\end{tabular} \\ \hline
\begin{tabular}[c]{@{}l@{}}\textit{\textbf{Learn Your Way}} \textgreater \\ \textit{\textbf{Textbook}}\end{tabular} &
  \begin{tabular}[l]{@{}l@{}}$Z$ = 1.89 \\ $p$ = .029\textsuperscript{*} \\ $r$ = .24\end{tabular} &
  \begin{tabular}[l]{@{}l@{}}$Z$ = 1.89 \\ $p$ = .03\textsuperscript{*}\\ $r$ = .25\end{tabular} \\

\end{tabular}
\end{table}

To ensure that performance outcomes were not driven by the duration of time students spent learning the content, we modeled the relationship between educational tool condition and learning duration on recall scores for both immediate and long-term recall assessments. Specifically, we employed a generalized additive model (GAM) to account for potential non-linear relationships between learning duration and assessment performance. 

For immediate recall scores, the model explained 13.4\% of the total deviance. The non-parametric smooth terms for learning duration were not significant for either \textit{Learn Your Way} ($edf=1$, $F=0.48$, $p=.49$) or \textit{Textbook} ($edf=1.82$, $F=1.43$, $p=.283$). Similar results were observed for long-term recall scores, where the model explained 9.1\% of the total deviance. Again, the smooth terms for learning duration were not significant for \textit{Learn Your Way} ($edf=1$, $F=1.53$, $p=.222$) or the \textit{Textbook} condition ($edf= 1$, $F= 0.15$, $p=.703$). Collectively, these findings suggest that learning duration did not meaningfully influence performance outcomes on either the immediate or long-term recall assessment, regardless of the experimental condition.

Finally, to account for variations in the retention interval across students, we employed another GAM to model the relationship between length of delay (3--7 days) and long-term recall performance. The model explained 8.2\% of the total deviance. The non-parametric smooth terms for delay duration were not significant for \textit{Learn Your Way} ($edf=1$, $F=1.09$, $p=.3$) or \textit{Textbook} ($edf=1$, $F=0.02$, $p=.903$), demonstrating that the specific length of delay did not significantly influence long-term recall within our required follow-up window of 3--7 days.

Regardless of experimental condition, assessment performance remained remarkably stable across time. Specifically, the \textit{Learn Your Way} participants achieved nearly identical percentage scores on the immediate recall assessment ($M = 9.2/12$; 77\%) and the long-term recall assessment ($M = 4.7/6$; 78\%). Similarly, \textit{Textbook} participants demonstrated a slight decrease 
of only one percentage point from immediate recall ($M = 8.2/12$; 68\%) to long-term recall ($M = 4/6$; 67\%). These results suggest that students retained the vast majority of acquired information, with negligible decay occurring across the 3--7 day retention interval. Consequently, the performance advantage observed for \textit{Learn Your Way} during the immediate assessment was maintained through the long-term recall phase.

As an exploratory analysis, we also investigated whether or not the use of more learning modalities, and thus, more content transformation capabilities, in the \textit{Learn Your Way} condition had an effect on students’ immediate and long-term recall performance. Table \ref{tab:table5} shows the different learning modalities that \textit{Learn Your Way} participants used during the learning phase. As a reminder, all participants landed in Immersive Text by default and 100\% of students utilized the QuizMe feature during their learning session.

\begin{table*}[t]
\centering
\caption{Number of \textit{Learn Your Way} participants who used one or more of the five core learning modalities provided in the tool.}
\vspace*{2mm} 
\label{tab:table5}
\renewcommand{\arraystretch}{1.5} 
\begin{tabular}{p{5.3cm}cc}
\toprule
\textbf{\begin{tabular}[c]{@{}l@{}}\textit{Learn Your Way}\\ Modality Usage \end{tabular}} &
  \begin{tabular}[c]{@{}c@{}}\textbf{Number of Participants}\\ (Total \textit{n} = 30)\end{tabular} \\ \hline
Immersive Text + QuizMe & 30 \\
Slides                  & 17 \\
Mindmap                 & 14 \\
Audio Lesson            & 6  \\
Video                   & 4  \\

\end{tabular}
\end{table*}

\begin{table*}[h]
\centering
\caption{\textit{Learn Your Way} modality usage across participants and mean scores for the immediate recall assessment ($n=30$; out of 12 points) and the long-term recall assessment ($n=29$; out of 6 points). 
\newline
\small \emph{*One of these participants did not complete the long-term recall assessment, hence the drop from total \textit{n}=30 during the immediate recall assessment to \textit{n}=29 for the long-term recall sample.}}
\vspace*{2mm} 
\label{tab:table6}
\renewcommand{\arraystretch}{1.5}

\begin{tabular}{p{5.3cm}ccc}
\toprule
\textbf{\begin{tabular}[c]{@{}l@{}}\textit{Learn Your Way}\\ Modality Usage \end{tabular}} &
\begin{tabular}[c]{@{}c@{}}\textbf{Number of Participants}\\ (Total \textit{n} = 30)\end{tabular} &
\begin{tabular}[c]{@{}c@{}}\textbf{Immediate Recall}\\ Mean (\textit{SD})\end{tabular} &
\begin{tabular}[c]{@{}c@{}}\textbf{Long-term Recall}\\ Mean (\textit{SD})\end{tabular} \\ \hline
Immersive Text only                   & 12  & 8.6 (1.8)  & 4.4 (1.3) \\
Immersive Text +1 modality & 4   & 10.8 (1.9) & 5.3 (0.5) \\
Immersive Text +2-4 modalities   & 14\textsuperscript{*} & 9.3 (2.4)  & 4.7 (1.3) \\ 
\end{tabular}

\end{table*}

We further split \textit{Learn Your Way} participants into students that used Immersive Text only, students that used Immersive Text plus one additional learning modality and students that used Immersive Text plus two or more learning modalities (Table \ref{tab:table6}). For the following analysis, we collapsed between the latter two groups when comparing immediate assessment scores between Immersive Text + QuizMe only ($n=12$; $Mdn=8.5$; $M=8.6$, $SD=1.8$) vs. Immersive Text + any additional modality(ies) ($n=18$; $Mdn=10$; $M=9.6$, $SD=2.3$). A two-sided Wilcoxon-Mann-Whitney test revealed no significant difference between groups ($Z=1.48$, $p=.144$, $r=.27$). Furthermore, while we see, directionally, that scores are higher for students that used more modalities in \textit{Learn Your Way}, we did not find a significant correlation between learning modality usage and performance on the immediate recall assessment (Pearson’s $r=.24$, $p=.206$). In other words, we did not find evidence that students who used more content transformation modalities had a better performance on the immediate assessment. This suggests that the pedagogical value of the system may reside in the quality of engagement with a chosen modality rather than the breadth of modalities accessed.

Similarly, we did not find a significant difference between Immersive Text + QuizMe only ($n=12$; $Mdn=4.5$; $M=4.4$, $SD=1.3$) vs. Immersive Text + any additional modality(ies) ($n=17$; $Mdn=5$; $M=4.8$, $SD=1.1$) for long-term recall scores ($Z=0.87$, $p=.396$, $r=.16$) nor did we find a significant correlation (Pearson’s $r=.17$, $p=.380$). 

For both exploratory analyses reported above, we acknowledge that statistical power to assess between-group differences and correlations was lower than our primary analyses, due to splitting our sample size for \textit{Learn Your Way} into two groups (see Section \ref{discussion} for more details).

\subsection{Learning Experience}


\begin{table*}[h]
\centering
\caption{Mean agreement ratings for learning experience statements shown by educational tool condition in middle two columns. Summary of directional Wilcoxon-Mann-Whitney tests shown in final column, including adjusted \textit{p}-values after multiple comparison corrections are applied. Asterisks denote significant differences at \textit{p}$<$ .05\textsuperscript{*} (or .01\textsuperscript{**} and .001\textsuperscript{***}). Effect sizes are reported as the rank-biserial correlation.}
\vspace*{2mm} 
\label{tab:table7}
\renewcommand{\arraystretch}{1.5}

\begin{tabular}{m{8.2cm}ccc} \toprule &

\multicolumn{2}{c}{\textbf{Agreement Ratings}} & \textbf{Test Statistics} \\[1.5ex]
\cline{2-4}
\textbf{Learning Experience Statements} & 
\begin{tabular}[c]{@{}c@{}}\textit{\textbf{Learn Your Way}}\\ Mean (\textit{SD})\end{tabular} & 
\begin{tabular}[c]{@{}c@{}}\textit{\textbf{Textbook}}\\ Mean (\textit{SD})\end{tabular} & 
\begin{tabular}[c]{@{}c@{}}\textit{Z}-score, \textit{p}-value,\\ \textit{r} effect size\end{tabular}\\  
\hline
    I felt like today's educational tool made me \newline more comfortable taking the assessment. &
    4.60 (0.50) & 3.83 (0.83) & 
\begin{tabular}[c]{@{}l@{}}$Z = 3.77$\\ $p_{adj}  < .001^{***}$\\ $r = .49$\end{tabular}\\

\hline
    The educational tool I used today would make me more effective at learning compared to other educational tools I currently use at home / school. & 
    4.17 (0.79) & 3.07 (0.98) & 
\begin{tabular}[c]{@{}l@{}}$Z = 4.11$\\ $p_{adj}  < .001^{***}$\\ $r = .53$\end{tabular}\\

\hline
    I would recommend this educational tool to \newline other students to support their learning needs. & 
    4.53 (0.78) & 3.97 (0.89) & 
\begin{tabular}[c]{@{}l@{}}$Z = 2.88$\\ $p_{adj} = .004^{**}$\\ $r = .37$\end{tabular}\\

\hline
    I found today's educational tool enjoyable to use. & 
    4.27 (0.64) & 3.67 (0.99) & 
\begin{tabular}[c]{@{}l@{}}$Z = 2.44$\\ $p_{adj} = .01^{*}$\\ $r = .31$\end{tabular}\\

\hline
    I would like to use today's educational tool to \newline support my learning needs in the future. & 
    4.43 (0.73) & 3.83 (1.05) & 
\begin{tabular}[c]{@{}l@{}}$Z = 2.42$\\ $p_{adj} = .01^{*}$\\ $r = .31$\end{tabular}\\

\hline
    I felt like today's educational tool helped me \newline gain a good understanding of the content. & 
    4.50 (0.57) & 4.07 (0.78) & 
\begin{tabular}[c]{@{}l@{}}$Z = 2.28$\\ $p_{adj} = .014^{*}$\\ $r = .29$\end{tabular}\\

\hline
    I felt like I performed well on the assessment. & 
    4.30 (0.70) & 3.83 (0.95) & 
\begin{tabular}[c]{@{}l@{}}$Z = 1.92$\\ $p_{adj} = .027^{*}$\\ $r = .25$\end{tabular}\\

\end{tabular}
\end{table*}


In support of our second hypothesis (H2), we observed meaningful differences when we analyzed students’ reported experience with the educational tools. Students in the \textit{Learn Your Way} condition ($Mdn=4$; $M=3.6$, $SD=0.6$) found the tool to be more useful for learning the educational content compared to \textit{Textbook} students ($Mdn=3$; $M=3.1$, $SD=0.7$). Specifically, a one-sided Wilcoxon-Mann-Whitney test revealed that \textit{Learn Your Way} participants rated the educational tool significantly higher on a 4-point Likert scale than \textit{Textbook} participants, before and after correcting for multiple comparisons with other experience metrics ($Z=3.08$, $p=.001$, $p_{adj}=.003$, $r=.40$). 

Furthermore, through a collection of agreement statements about the educational tool, we can see how the \textit{Learn Your Way} platform positively impacted students’ perceived learning experience as compared to students in the \textit{Textbook} condition (as shown in Table \ref{tab:table7}). Specifically, students in the \textit{Learn Your Way} condition reported higher levels of agreement compared to students in the \textit{Textbook} condition across all of our experiential measures. Directional Wilcoxon-Mann-Whitney tests were conducted on each statement's agreement ratings, followed by FDR corrections for multiple comparisons. Table \ref{tab:table7} provides a full breakdown of the included statements and mean values by educational tool condition, as well as the associated test statistics with the adjusted \textit{p}-values reported. The most significant differences revealed between the \textit{Learn Your Way} platform and the \textit{Textbook} condition suggest \textit{Learn Your Way} has the strongest influence on students' comfort in test-taking, and their perception of the tool's effectiveness for learning compared to other educational tools already being used at home or in their classroom.


\section{Qualitative Findings}
Data from the qualitative interviews shed light on the trends we observed in the quantitative data. 

\subsection{RQ1 and RQ2: Learning Outcomes}
\subsubsection{Dynamic representations of content can effectively help students learn.}
One of the unique attributes of \textit{Learn Your Way} is that it provides multiple interactive transformations in one learning experience. This approach is based on established learning principles, such as Mayer and Moreno’s “Cognitive Theory of Multimedia Learning” \citep{mayer1998cognitive} and Paivio’s “Mental representations: A dual coding approach,” \citep{Paivio1990} which posit that encoding information in multiple formats can make learning more effective by engaging multiple senses and cognitive channels, and by simplifying complex ideas. Students in the present study expressed value in the ability to access multiple learning modalities. These were perceived as useful for clarifying concepts and improving comprehension and retention of the learning material: 

\begin{quote}
    “Reading it multiple times but in different ways made it stick in my mind. In the assessment, some topics I remembered from the slide, some I remembered from reading. So, reading it multiple times but in different ways made it stick in my mind.” 
    \newline - P86, Urban, \textit{Learn Your Way}
\end{quote}

\begin{quote}
    “If I didn’t feel I was getting good information from one [modality], I could go to the next one… Immersive Text and the slides are really good for a study tool especially. Mindmap helped me put everything together and really see where everything fits in here… I think because it’s the same information, you just kind of can play around with it and go to different things like Slides and stuff. I think that, in itself, helped me to remember [the material] and see what’s really going on.” 
    \newline - P20, Suburban, \textit{Learn Your Way}
\end{quote}

Conversely, our qualitative feedback revealed that participants from the \textit{Textbook} condition desired the ability to view content in multiple formats. Despite being unaware of the \textit{Learn Your Way} condition, several \textit{Textbook} participants described features already included in \textit{Learn Your Way} when asked what their ideal learning tool would look like:
\begin{quote}
    “I’m more of a visual learner, so if I see something visual, I better understand it than words or text.”  
    \newline - P09, Urban, \textit{Textbook}
\end{quote}

Students in both conditions highly valued and preferred interactive and multimodal learning materials, as evidenced by their feedback. This approach felt like an important way to create effective and engaging learning experiences. Instead of reading passively through lengthy passages, students using \textit{Learn Your Way} absorbed content in diverse ways that aligned with their preferences, which was perceived to facilitate a deeper understanding and better retention of the content.

\subsubsection{Summaries and smaller chunks of text via Immersive Text in \textit{Learn Your Way} improved students’ comprehension of the content}

A key feature of the Immersive Text modality in \textit{Learn Your Way} was that it broke down the chapter into manageable chunks of information, and generated interactive practice and quizzes to test the level of comprehension of each section. Given that participants in the experimental condition spent most of their time utilizing this feature, its effectiveness likely played a role in the learning outcomes we observed. Many students spoke about the ways Immersive Text helped them digest and learn the content:

\begin{quote}
    “I do feel the way that they structured it - having only a few paragraphs or split up text definitely helps my brain not get overwhelmed by the whole thing. So not just putting a whole chunk of text that is relating to the whole topic and then just being like, ‘okay here.’"
    \newline - P40, Suburban, \textit{Learn Your Way}
\end{quote}

\begin{quote}
    “Like anything in my life, if I see something I have to tackle…and it’s one big thing - I struggle breaking it up into smaller parts. If it’s reading a really long article for a class or cleaning my room, I look at it and I’m like, ‘whoa, that’s a lot of stuff. I can’t do it.’  If it’s broken up for me, I feel less overwhelmed by it.” 
    \newline - P13, Suburban, \textit{Learn Your Way}
\end{quote}

In contrast, some \textit{Textbook} students felt overwhelmed by having only the chapter’s content, which made it harder to focus and retain information. We also observed that some students in this condition attempted to create summaries and break the content up on their own by using the annotation tools within the digital textbook.

\begin{quote}
    “Because even though the paragraphs aren't that big, there's still enough paragraphs where you'll forget your information.” 
    \newline - P54, Rural, \textit{Textbook}
\end{quote}

\begin{quote}
    “Maybe if it stated the key points a little better because only about half of the article has like an actual head headliner, you know, forgot what they're called…so yeah, I mean maybe just like breaking down the big chunks of text a little better.” 
    \newline -P94, Suburban, \textit{Textbook}
\end{quote}

In summary, the qualitative feedback from students in both conditions reveal a preference for content presented in more manageable segments. In the absence of these summarized chunks of information, students expressed difficulty retaining the information from the textbook chapter, which likely impacted their ability to recall the information. This highlights the need to consider the impact of visual presentation and organization in the design of content transformation tools.

\subsubsection{The value of assessments via QuizMe was a key driver in performance outcomes}
One of the helpful ways students effectively learned the material in \textit{Learn Your Way} was through the QuizMe feature. All participants in \textit{Learn Your Way} acknowledged the role of quizzes and feedback on their performance. This is consistent with previous research that demonstrates that quizzes can significantly enhance learning outcomes by offering additional practice exercises in areas where a student is struggling \citep{ChangChien2024}. This approach helps learners attain mastery at their own pace and speeds up the learning process \citep{al2022perceptions, Kulik2016}. Quizzes were a favored way to test comprehension of the material, and feedback instilled confidence in learning by increasing perceptions of knowledge retention for students in the \textit{Learn Your Way} condition. 

\begin{quote}
    “The quizzes for sure helped me get the right answers. It helped me even if I needed help, it gave a hint if I didn't understand it. I feel like it also made it really easy to explain or understand.” 
    \newline - P20, Urban, \textit{Learn Your Way}
\end{quote}

\begin{quote}
    “I liked how it gave me feedback and told me why I got it right versus just bringing me to the next question right away. It was helpful and made me realize how my thought process was and how I got to the answer.” 
    \newline - P11, Suburban, \textit{Learn Your Way}
\end{quote}

In contrast, students in the \textit{Textbook} condition reported missing the ability to check their understanding and receive feedback through guided activities and tools such as  quizzes, flashcards and look up tools:

\begin{quote}
    “I would probably make a tool where everything is together, like the test taking with the text so I can select what I want to learn more about in the text and put it in a test where I’m able to answer the questions if I know it goes away, if I don’t I have to keep studying it and learn more.” 
    \newline - P12, Suburban, \textit{Textbook} 
\end{quote}

\begin{quote}
    “Like the ideal learning tool, like if I were to learn this article a different way, I mean, yeah, like I said, I mean flashcards I find useful. I definitely think once you read it, once you like read a certain amount of it or whatever, like certain checkpoints I guess it could quiz you on what you've done so far to review to make sure that you got all the key points.”
    \newline - P94, Suburban, \textit{Textbook}
\end{quote}

Feedback from participants aligns with educational theories about active engagement in learning and emphasizes the need to provide active checkpoints in learning experiences \citep{Moreno2000}. This approach allows students to leverage metacognitive strategies to reflect on their understanding of the learning material and increase the likelihood of retaining information \citep{Zimmerman2009}.

\subsection{RQ3: Learning Experiences}
\subsubsection{The variety of capabilities and self-directed nature of \textit{Learn Your Way} played a critical role in creating a positive learning experience}

Access to multiple transformations and the freedom to choose which ones to learn with resulted in an engaging learning experience for students in the \textit{Learn Your Way} condition. This is also consistent with research that advocates for learners to play a central role in their own learning journeys \citep{Deci2000, Zimmerman2009, zimmerman2000}.

\begin{quote}
    “I personally like going in my own structured way… It’s just a learning preference… I like it more of me leading it because I feel more in control of my learning…it allows me to go my own pace and let me learn how I want to learn.” 
    \newline - P08, Urban, \textit{Learn Your Way}
\end{quote}

\begin{quote}
    “I like how many options it has for different ways of taking in the information because I definitely like learning it one way and then being able to pick another way to review…”
    \newline - P13, Suburban, \textit{Learn Your Way}
\end{quote}

Conversely, students in the \textit{Textbook} condition who did not have access to transformations, noted the challenges of this learning experience and desired more interactions. 

\begin{quote}
    “I guess I liked how it was kind of just all there and I could read it. What I didn't like…it was just like kind of boring reading it… I was just like sitting there trying to memorize it. Reading it…I usually learn best like videos and stuff. I can’t like read stuff and store in memory… I guess I'm like more of like a visual learner.” 
    \newline - P19, Suburban, \textit{Textbook}
\end{quote}

Feedback from students in both conditions demonstrates the opportunity to facilitate student-led learning through the provision of multiple content transformations within a single learning experience.

\subsubsection{Auditory transformations were perceived as useful for combining learning with other activities}

While a small percentage (\textit{n}= 6, 20\%) of \textit{Learn Your Way} participants used Audio or Video modalities, several of them recognized it as a way to multitask and incorporate learning into their daily routines. Because audio content does not require much physical interaction, students liked the idea of completing less cognitively demanding tasks while listening to educational content. They explained that it would help to use their time more efficiently and help them to feel more productive while completing routine tasks.

\begin{quote}
    “I do listen to like a lot of podcasts in general, but I always like to have like, stuff playing while I'm doing stuff. So like, like I was saying before, like, oh, if you're like cooking or cleaning your room or just doing anything really like, I don't know, like even like hobbies, like you're like drawing or something. Like just having, I always like to have something playing in the background. So like when I have the option to have that for like school stuff, I just like have, I feel it feels more productive. Like I can get it done while I'm doing other stuff, while like retaining the information as well.” 
    \newline - P86, Urban, \textit{Learn Your Way}
\end{quote}

\begin{quote}
    “I mean it's the same with the Video I guess. It's like, like people talking rather than somebody explaining they're, I mean, in my mind pretty similar in use case. Like if I'm driving or like, you know, cleaning or doing something where I can't like read and I can just throw this on and learn about it. That would be the use case.”
    \newline - P01, Urban, \textit{Learn Your Way}
\end{quote}

The \textit{Textbook} condition, on the other hand, required more visual focus to process the content in the textbook chapter. Without access to text-to-speech tools or additional support, multitasking within the digital textbook would therefore be challenging. It is worth noting however that there are open questions about the impact of multitasking on the comprehension and recall of information and which tasks are suitable for this \citep{Coens2011}. It is therefore crucial to consider approaches to support metacognitive strategies (e.g. reflection) with auditory modalities to ensure effective learning.

\section{Discussion} \label{discussion}

Our results are in support of our hypotheses, which demonstrate that \textit{Learn Your Way} can lead to both improved learning outcomes and experiences. AI-powered learning platforms can be a powerful driver of pedagogical innovation. In particular, our experiment reveals that these tools are not merely supplementary aids but can play a critical role in reshaping how students interact with and comprehend academic content. While many studies exploring the impact of AI tools on educational outcomes have been mixed (e.g.,\citep{do2025, Sun2024}), our research demonstrates that \textit{Learn Your Way} can enhance students' immediate and long-term recall of information and create a more effective learning environment compared to a traditional textbook. 

The experimental results suggest that this can be attributed to several factors. First, learning is a complex and multifaceted process and students exhibit diverse preferences in how they process information and acquire knowledge, including visual, kinesthetic, and reading/writing approaches  \citep{mccall2024future}. Traditional instructional methods often fail to address these unique needs at scale, leading to suboptimal educational experiences \citep{Villegas2024}. Instead, engaging with information through multiple representational formats, such as slide decks, audio lessons, or visual mind-maps, facilitates the construction of a more robust and comprehensive mental model of the content \citep{AINSWORTH1999}. This is because diverse representations can clarify complex ideas and highlight distinct facets of a concept. \textit{Learn Your Way} addresses this need by offering students the ability to select their preferred learning modalities, enabling them to explore concepts from a range of engaging and diverse formats. Second, simplified and multimodal presentations can reinforce learning, leading to greater retention when information is seen, heard, and actively engaged with \citep{clark1991dual}. By breaking down content into manageable chunks and allowing students to personalize the presentation of the content in dynamic formats based on their learning preferences,  we believe students forged better mental connections between different representations, resulting in better comprehension of the content compared to students who only engaged with a digital textbook.  

Drawing from established literature on the benefits of adaptive quizzing and personalized feedback \citep{frank2024influence}, this study’s quantitative and qualitative findings underscore their critical role in the learning experience. All students who used \textit{Learn Your Way} experienced quizzing and feedback. Participants' qualitative feedback further highlighted the value of these capabilities, with many emphasizing how quizzes aided in testing comprehension and retention, while the associated feedback was instrumental in helping them understand the rationale behind their answers.

Given the power of content transformation capabilities on student learning, one might wonder why there was no direct correlation between engagement with usage of more content transformation modalities and better learning outcomes among students who used \textit{Learn Your Way}. This analysis was exploratory and the pattern of data is in the direction we expected (i.e., more use of \textit{Learn Your Way} modalities was correlated with better learning outcomes) but not statistically significant.  We expect this may have occurred for two reasons. First, this analysis required us to segment our \textit{Learn Your Way} students into two different groups, which significantly reduced our sample size and therefore power to detect statistical significance. Second, our qualitative feedback suggests that the usage of additional modalities, such as Slides and Video were often treated as supplementary resources for content review, following primary engagement with the Immersive Text and QuizMe features.

\begin{quote}
    “I feel like [Slides] would be good for review after I already read the Immersive Text if I wanted to review the main points.” - P72, Rural, \textit{Learn Your Way}
\end{quote}

Our findings therefore suggest two potential explanations for the observed phenomena. First, a larger sample size could provide a more robust basis for investigating whether increased use of content transformation capabilities positively impacts learning outcomes. Second, it may be the case that a greater quantity of engagement with content transformation features does not, in itself, lead to improved learning outcomes. Rather, our results corroborate previous research demonstrating that enhanced learning is contingent upon the strategic alignment of an individual's learning preferences with the specific demands of the task \citep{Cordova1996, schneider2018autonomy}. In other words, the observed benefits are likely driven by the availability of multimodal options and the resulting learner agency, rather than a cumulative effect of exposure to multiple formats. As our current design did not manipulate modality exposure experimentally, we refrain from implying that using a higher number of transformations leads to superior outcomes. Instead, the results suggest that providing a diverse ``menu" of representations allows students to select the single most effective format for their individual needs. While these tools offer significant value, their effectiveness is maximized when learners have the autonomy to select the specific transformations that align with their personal learning goals and contextual needs. Crucially, our data indicates that effectiveness is not predicated on the number of modalities utilized, but rather on the student’s ability to find a representation that aligns with their specific learning needs. 

Furthermore, it is important to note that \textit{Learn Your Way} is not predicated on the efficacy of a single content representation, but rather on the theoretical synergy between learner agency and the benefits of multiple representations (as grounded in Dual Coding Theory and Multimedia Learning principles). Therefore, attempting to isolate the performance contribution of any individual content transformation from the element of student choice would fundamentally detach the intervention from its core theoretical foundation, misrepresenting the mechanism by which we hypothesize the system promotes positive performance outcomes.

Our results also demonstrate that students who engaged with \textit{Learn Your Way} reported a more positive learning experience compared to those who used a traditional textbook. The present study's findings are supported by a body of literature suggesting that transforming educational content into multiple representations can make the learning process more enjoyable, engaging, and motivating \citep{AINSWORTH1999}. This is particularly powerful when students are able to select representations that best match their individual learning preferences \citep{surjono2015effects,Popescu2010}. \textit{Learn Your Way}, which is capable of dynamically transforming content, represents a critical evolution from standard digital textbooks by enabling more tailored and engaging educational experiences for students. 

Finally, our qualitative feedback indicated a distinct preference for content transformation tools over other generative AI applications, citing their substantial contribution to the learning experience. This finding underscores a potential differentiation in the pedagogical impact of various AI tools, where those that enable students to explore concepts from multiple perspectives may offer a more direct and perceived benefit to student learning.

\subsection{Limitations and Future Work}
While our findings provide evidence that AI-powered content transformation tools can positively impact learning outcomes and experiences, our study is not without limitations. First, our study took place in an experimental lab setting. Future research should explore best practices for how and when to utilize these content transformation tools within a classroom environment, and collect longitudinal outcomes on learning and retention. Second, a limitation of the current study design is the lack of a feature-matched control group. While the \textit{Textbook} condition represents the current standard for digital learning, it lacks the interactive scaffolds (e.g., quizzes and feedback) present in \textit{Learn Your Way}. Consequently, the observed gains may be attributed to the combined effect of content transformation and these interactive elements. Future research should employ a control condition that includes non-AI-generated quizzes to isolate the specific impact of the generative content transformations. Third, while the observed effect sizes for immediate ($r = .24$) and long-term recall ($r = .25$) are characterized as small to medium by traditional conventions, their practical significance in a classroom setting is noteworthy. 

In educational contexts, small improvements in recall can lead to significant cumulative gains in student mastery over a full academic year. Furthermore, these effects were achieved with a relatively brief intervention (up to 40 minutes). For a teacher, moving a student from an average score of $8.2/12$ to $9.2/12$ -- as seen in our immediate recall results -- represents the difference between a ``D" and a ``C" grade. Given that \textit{Learn Your Way} is an automated, AI-driven tool, these gains suggest a scalable method for providing personalized support that would otherwise require intensive one-on-one tutoring. Thus, the tool demonstrates a ``low-effort, high-yield" potential for augmenting standard classroom instruction. 

Finally, while this study provides a foundational proof-of-concept, several factors present potential threats to its external validity. Specifically, the controlled lab environment and the focus on a single age group (high school students) and a single subject area (neuroscience) limit the generalizability of our findings to more ecologically valid settings. To address this, future research should transition from the lab to a longitudinal field study within diverse classroom environments. This will allow researchers to observe how the efficacy of AI-powered learning persists across various disciplines -- such as mathematics or literature -- where the nature of ``transformation" may differ structurally. Furthermore, by expanding the participant pool to include middle school and university students, researchers can identify the specific developmental stages at which these interactive representations provide the most significant pedagogical benefit.

\section{Conclusion}
Our experiment results demonstrate the potential of using generative AI to transform traditional textbook chapters into more engaging formats. We found that US high school students from the Chicago area preferred \textit{Learn Your Way} over a traditional digital textbook, appreciating the dynamic representations of the content, quizzes, and feedback. The burgeoning field of content transformation tools offers a compelling opportunity to not only enhance learning outcomes but also to enrich the overall student experience.


\section*{Conflict of Interest Statement}
Courtney Heldreth, Diana Akrong, Laura Vardoulakis, Yael Haramaty, Lidon Hackmon, and Lior Belinsky are employees of Google LLC, the developer of \textit{Learn Your Way} discussed in this study. This study was funded by Google LLC. Google LLC  was involved in the study design, collection, analysis, interpretation of data, the writing of this article, and the decision to submit it for publication. The authors declare that this commercial affiliation did not influence the objectivity of the research nor the decision to publish.

\section*{Author Contributions}
\begin{itemize}
    \item []\textit{Study concept and methodological design:} Akrong, Heldreth, Vardoulakis
    \item []\textit{Data collection:} Akrong, Heldreth, Miller
    \item []\textit{Technical software support during the study:} Belinsky, Hackmon
    \item []\textit{Data analysis:} Akrong, Heldreth, Miller, Siebert, Tapia, Tootill
    \item []\textit{Drafting of the original manuscript:} Akrong, Heldreth, Miller, Vardoulakis
    \item []\textit{Critical revisions of the manuscript:} All authors
    \item []\textit{Obtained funding:} Haramaty, Heldreth
    \item []\textit{Study supervision:} Heldreth
\end{itemize}

\section*{Acknowledgments}
We’d like to thank the Google Research Education Team for building \textit{Learn Your Way} and for all of the support they provided during the study: Alicia Martín, Amir Globerson,  Amy Wang, Anirudh Shekawat, Anisha Choudhury, Anna Iurchenko, Avinatan Hassidim, Ayça Çakmakl,  Ayelet Shasha Evron, Charlie Yang, Gal Elidan, Hairong Mu, Ian Li, Ido Cohen, Katherine Chou, Komal Singh, Lev Borovoi, Michael Fink, Niv Efron, Preeti Singh, Rena Levitt, Shashank Agarwal, Shay Sharon, Tracey Lee-Joe, Xiaohong Hao, Yael Gold-Zamir, Yishay Mor, Yoav Bar Sinai, and Yossi Matias. Thank you to the Bold Insight team and Gavin Lew for their instrumental support throughout the study. We also thank Anna Iurchenko for her assistance with the figures in this paper, Adam Mansour, Julia Wilkowski, Beverly Freeman, and Roma Ruparel for their feedback on the study design, and Amy Keeling for her study support.  Finally, we thank all of the study participants for their willingness to share their experiences and contribute to our research.

\bibliographystyle{Frontiers_LaTeX_Templates/Frontiers-Harvard} 
\bibliography{main-frontiers}

@String{Computing = "Computing" }

@String{Computer = "{IEEE} Computer" }

@String{Academic = "Academic Press" }

@String{Springer = "Springer-Verlag" }

@inproceedings{stasko1993,
author = {Stasko, John and Badre, Albert and Lewis, Clayton},
title = {Do algorithm animations assist learning? an empirical study and analysis},
year = {1993},
isbn = {0897915755},
publisher = {Association for Computing Machinery},
address = {New York, NY, USA},
url = {https://doi.org/10.1145/169059.169078},
doi = {10.1145/169059.169078},
booktitle = {Proceedings of the INTERACT '93 and CHI '93 Conference on Human Factors in Computing Systems},
pages = {61–66},
numpages = {6},
location = {Amsterdam, The Netherlands},
series = {CHI '93}
}

@inproceedings{russell1993,
author = {Russell, Daniel M. and Stefik, Mark J. and Pirolli, Peter and Card, Stuart K.},
title = {The cost structure of sensemaking},
year = {1993},
isbn = {0897915755},
publisher = {Association for Computing Machinery},
address = {New York, NY, USA},
url = {https://doi.org/10.1145/169059.169209},
doi = {10.1145/169059.169209},
booktitle = {Proceedings of the INTERACT '93 and CHI '93 Conference on Human Factors in Computing Systems},
pages = {269–276},
numpages = {8},
location = {Amsterdam, The Netherlands},
series = {CHI '93}
}

@inproceedings{abowd1998,
author = {Abowd, Gregory D. and Atkeson, Christopher G. and Brotherton, Jason and Enqvist, Tommy and Gulley, Paul and LeMon, Johan},
title = {Investigating the capture, integration and access problem of ubiquitous computing in an educational setting},
year = {1998},
isbn = {0201309874},
publisher = {ACM Press/Addison-Wesley Publishing Co.},
address = {USA},
url = {https://doi.org/10.1145/274644.274704},
doi = {10.1145/274644.274704},
booktitle = {Proceedings of the SIGCHI Conference on Human Factors in Computing Systems},
pages = {440–447},
numpages = {8},
location = {Los Angeles, California, USA},
series = {CHI '98}
}

@inproceedings{scaife1997,
author = {Scaife, Michael and Rogers, Yvonne and Aldrich, Frances and Davies, Matt},
title = {Designing for or designing with? Informant design for interactive learning environments},
year = {1997},
isbn = {0897918029},
publisher = {Association for Computing Machinery},
address = {New York, NY, USA},
url = {https://doi.org/10.1145/258549.258789},
doi = {10.1145/258549.258789},
booktitle = {Proceedings of the ACM SIGCHI Conference on Human Factors in Computing Systems},
pages = {343–350},
numpages = {8},
location = {Atlanta, Georgia, USA},
series = {CHI '97}
}

@inproceedings{zuckerman2005,
author = {Zuckerman, Oren and Arida, Saeed and Resnick, Mitchel},
title = {Extending tangible interfaces for education: digital montessori-inspired manipulatives},
year = {2005},
isbn = {1581139985},
publisher = {Association for Computing Machinery},
address = {New York, NY, USA},
url = {https://doi.org/10.1145/1054972.1055093},
doi = {10.1145/1054972.1055093},
booktitle = {Proceedings of the SIGCHI Conference on Human Factors in Computing Systems},
pages = {859–868},
numpages = {10},
location = {Portland, Oregon, USA},
series = {CHI '05}
}

@inproceedings{Saerbeck2010,
author = {Saerbeck, Martin and Schut, Tom and Bartneck, Christoph and Janse, Maddy D.},
title = {Expressive robots in education: varying the degree of social supportive behavior of a robotic tutor},
year = {2010},
isbn = {9781605589299},
publisher = {Association for Computing Machinery},
address = {New York, NY, USA},
url = {https://doi.org/10.1145/1753326.1753567},
doi = {10.1145/1753326.1753567},
booktitle = {Proceedings of the SIGCHI Conference on Human Factors in Computing Systems},
pages = {1613–1622},
numpages = {10},
location = {Atlanta, Georgia, USA},
series = {CHI '10}
}

@inproceedings{winkler2020,
author = {Winkler, Rainer and Hobert, Sebastian and Salovaara, Antti and S\"{o}llner, Matthias and Leimeister, Jan Marco},
title = {Sara, the Lecturer: Improving Learning in Online Education with a Scaffolding-Based Conversational Agent},
year = {2020},
isbn = {9781450367080},
publisher = {Association for Computing Machinery},
address = {New York, NY, USA},
url = {https://doi.org/10.1145/3313831.3376781},
doi = {10.1145/3313831.3376781},
booktitle = {Proceedings of the 2020 CHI Conference on Human Factors in Computing Systems},
pages = {1–14},
numpages = {14},
location = {Honolulu, HI, USA},
series = {CHI '20}
}

@inproceedings{Leong2024,
author = {Leong, Joanne and Pataranutaporn, Pat and Danry, Valdemar and Perteneder, Florian and Mao, Yaoli and Maes, Pattie},
title = {Putting Things into Context: Generative AI-Enabled Context Personalization for Vocabulary Learning Improves Learning Motivation},
year = {2024},
isbn = {9798400703300},
publisher = {Association for Computing Machinery},
address = {New York, NY, USA},
url = {https://doi.org/10.1145/3613904.3642393},
doi = {10.1145/3613904.3642393},
booktitle = {Proceedings of the 2024 CHI Conference on Human Factors in Computing Systems},
articleno = {677},
numpages = {15},
location = {Honolulu, HI, USA},
series = {CHI '24}
}

@inproceedings{kumar2019,
author = {Kumar, Priya C. and Chetty, Marshini and Clegg, Tamara L. and Vitak, Jessica},
title = {Privacy and Security Considerations For Digital Technology Use in Elementary Schools},
year = {2019},
isbn = {9781450359702},
publisher = {Association for Computing Machinery},
address = {New York, NY, USA},
url = {https://doi.org/10.1145/3290605.3300537},
doi = {10.1145/3290605.3300537},
booktitle = {Proceedings of the 2019 CHI Conference on Human Factors in Computing Systems},
pages = {1–13},
numpages = {13},
location = {Glasgow, Scotland Uk},
series = {CHI '19}
}

@article{Reber2009,
author = {Rolf Reber and Hilde Hetland and Weiqin Chen and Elisabeth Norman and Therese Kobbeltvedt},
title = {Effects of Example Choice on Interest, Control, and Learning},
journal = {Journal of the Learning Sciences},
volume = {18},
number = {4},
pages = {509--548},
year = {2009},
publisher = {Routledge},
doi = {10.1080/10508400903191896},


URL = { 
    
        https://doi.org/10.1080/10508400903191896
    
    

},
eprint = { 
    
        https://doi.org/10.1080/10508400903191896
    
    

}

}

@article{Johnson2022,
author = {Johnson, Jazette and Arnold, Vitica and Piper, Anne Marie and Hayes, Gillian R.},
title = {"It's a lonely disease": Cultivating Online Spaces for Social Support among People Living with Dementia and Dementia Caregivers},
year = {2022},
issue_date = {November 2022},
publisher = {Association for Computing Machinery},
address = {New York, NY, USA},
volume = {6},
number = {CSCW2},
url = {https://doi.org/10.1145/3555133},
doi = {10.1145/3555133},
journal = {Proc. ACM Hum.-Comput. Interact.},
month = nov,
articleno = {408},
numpages = {27}
}

@inproceedings{Wong2025,
author = {Wong, Novia and Cheng, Nai-Yu and Oewel, Bruna and Genuario, Katherine E and Stoeckl, SarahElizabeth and Schueller, Stephen M. and Ahmed, Iftekhar and van der Hoek, Andr\'{e} and Reddy, Madhu},
title = { 'It's a spectrum': Exploring Autonomy, Competence, and Relatedness in Software Development Processes and Tools},
year = {2025},
isbn = {9798400713941},
publisher = {Association for Computing Machinery},
address = {New York, NY, USA},
url = {https://doi.org/10.1145/3706598.3713250},
doi = {10.1145/3706598.3713250},
booktitle = {Proceedings of the 2025 CHI Conference on Human Factors in Computing Systems},
articleno = {151},
numpages = {19},
location = {
},
series = {CHI '25}
}

@inproceedings{Chanenson2023,
author = {Chanenson, Jake and Sloane, Brandon and Rajan, Navaneeth and Morril, Amy and Chee, Jason and Huang, Danny Yuxing and Chetty, Marshini},
title = {Uncovering Privacy and Security Challenges In K-12 Schools},
year = {2023},
isbn = {9781450394215},
publisher = {Association for Computing Machinery},
address = {New York, NY, USA},
url = {https://doi.org/10.1145/3544548.3580777},
doi = {10.1145/3544548.3580777},
booktitle = {Proceedings of the 2023 CHI Conference on Human Factors in Computing Systems},
articleno = {592},
numpages = {28},
location = {Hamburg, Germany},
series = {CHI '23}
}

@inproceedings{chang2024,
author = {Chang, Michael Alan and Wong, Richmond Y. and Breideband, Thomas and Philip, Thomas M. and McKoy, Ashieda and Cortez, Arturo and D'Mello, Sidney K.},
title = {Co-design Partners as Transformative Learners: Imagining Ideal Technology for Schools by Centering Speculative Relationships},
year = {2024},
isbn = {9798400703300},
publisher = {Association for Computing Machinery},
address = {New York, NY, USA},
url = {https://doi.org/10.1145/3613904.3642559},
doi = {10.1145/3613904.3642559},
booktitle = {Proceedings of the 2024 CHI Conference on Human Factors in Computing Systems},
articleno = {566},
numpages = {15},
location = {Honolulu, HI, USA},
series = {CHI '24}
}

@inproceedings{nicholson2022,
author = {Nicholson, Rebecca and Bartindale, Tom and Kharrufa, Ahmed and Kirk, David and Walker-Gleaves, Caroline},
title = {Participatory Design Goes to School: Co-Teaching as a Form of Co-Design for Educational Technology},
year = {2022},
isbn = {9781450391573},
publisher = {Association for Computing Machinery},
address = {New York, NY, USA},
url = {https://doi.org/10.1145/3491102.3517667},
doi = {10.1145/3491102.3517667},
booktitle = {Proceedings of the 2022 CHI Conference on Human Factors in Computing Systems},
articleno = {150},
numpages = {17},
location = {New Orleans, LA, USA},
series = {CHI '22}
}

@inproceedings{Melfi2020,
author = {Melfi, Giuseppe and M\"{u}ller, Karin and Schwarz, Thorsten and Jaworek, Gerhard and Stiefelhagen, Rainer},
title = {Understanding what you feel: A Mobile Audio-Tactile System for Graphics Used at Schools with Students with Visual Impairment},
year = {2020},
isbn = {9781450367080},
publisher = {Association for Computing Machinery},
address = {New York, NY, USA},
url = {https://doi.org/10.1145/3313831.3376508},
doi = {10.1145/3313831.3376508},
booktitle = {Proceedings of the 2020 CHI Conference on Human Factors in Computing Systems},
pages = {1–12},
numpages = {12},
location = {Honolulu, HI, USA},
series = {CHI '20}
}

@inproceedings{Liu2024,
author = {Liu, Ziyi and Zhu, Zhengzhe and Zhu, Lijun and Jiang, Enze and Hu, Xiyun and Peppler, Kylie A and Ramani, Karthik},
title = {ClassMeta: Designing Interactive Virtual Classmate to Promote VR Classroom Participation},
year = {2024},
isbn = {9798400703300},
publisher = {Association for Computing Machinery},
address = {New York, NY, USA},
url = {https://doi.org/10.1145/3613904.3642947},
doi = {10.1145/3613904.3642947},
booktitle = {Proceedings of the 2024 CHI Conference on Human Factors in Computing Systems},
articleno = {659},
numpages = {17},
location = {Honolulu, HI, USA},
series = {CHI '24}
}

@inproceedings{Chen2024,
author = {Chen, Liuqing and Xiao, Shuhong and Chen, Yunnong and Song, Yaxuan and Wu, Ruoyu and Sun, Lingyun},
title = {ChatScratch: An AI-Augmented System Toward Autonomous Visual Programming Learning for Children Aged 6-12},
year = {2024},
isbn = {9798400703300},
publisher = {Association for Computing Machinery},
address = {New York, NY, USA},
url = {https://doi.org/10.1145/3613904.3642229},
doi = {10.1145/3613904.3642229},
booktitle = {Proceedings of the 2024 CHI Conference on Human Factors in Computing Systems},
articleno = {649},
numpages = {19},
location = {Honolulu, HI, USA},
series = {CHI '24}
}

@article{AINSWORTH1999,
title = {The functions of multiple representations},
journal = {Computers \& Education},
volume = {33},
number = {2},
pages = {131-152},
year = {1999},
issn = {0360-1315},
doi = {https://doi.org/10.1016/S0360-1315(99)00029-9},
url = {https://www.sciencedirect.com/science/article/pii/S0360131599000299},
author = {Shaaron Ainsworth}
}

@incollection{mayer2002,
title = {Multimedia learning},
  booktitle={Psychology of learning and motivation},
publisher = {Academic Press},
volume = {41},
pages = {85-139},
year = {2002},
issn = {0079-7421},
doi = {https://doi.org/10.1016/S0079-7421(02)80005-6},
url = {https://www.sciencedirect.com/science/article/pii/S0079742102800056},
author = {Richard E. Mayer}
}

@article{mayer1998cognitive,
  title={A cognitive theory of multimedia learning: Implications for design principles},
  author={Mayer, Richard E and Moreno, Roxana},
  journal={Journal of educational psychology},
  volume={91},
  number={2},
  pages={358--368},
  year={1998}
}

@article{anderson1985,
author = {John R. Anderson  and C. Franklin Boyle  and Brian J. Reiser },
title = {Intelligent Tutoring Systems},
journal = {Science},
volume = {228},
number = {4698},
pages = {456-462},
year = {1985},
doi = {10.1126/science.228.4698.456},
URL = {https://www.science.org/doi/abs/10.1126/science.228.4698.456},
eprint = {https://www.science.org/doi/pdf/10.1126/science.228.4698.456}}

@incollection{corbett1997,
title = {Chapter 37 - Intelligent Tutoring Systems},
editor = {Marting G. Helander and Thomas K. Landauer and Prasad V. Prabhu},
booktitle = {Handbook of Human-Computer Interaction (Second Edition)},
publisher = {North-Holland},
edition = {Second Edition},
address = {Amsterdam},
pages = {849-874},
year = {1997},
isbn = {978-0-444-81862-1},
doi = {https://doi.org/10.1016/B978-044481862-1.50103-5},
url = {https://www.sciencedirect.com/science/article/pii/B9780444818621501035},
author = {Albert T. Corbett and Kenneth R. Koedinger and John R. Anderson}
}

@article{graesser2004autotutor,
  title={AutoTutor: A tutor with dialogue in natural language},
  author={Graesser, Arthur C and Lu, Shulan and Jackson, George Tanner and Mitchell, Heather Hite and Ventura, Mathew and Olney, Andrew and Louwerse, Max M},
  journal={Behavior Research Methods, Instruments, \& Computers},
  volume={36},
  number={2},
  pages={180--192},
  year={2004},
  publisher={Springer}
}

@article{tu2025empowering,
  title={Empowering personalized learning with generative artificial intelligence: Mechanisms, challenges and pathways},
  author={Tu, Yaxin and Chen, Jili and Huang, Changqin},
  journal={Frontiers of Digital Education},
  volume={2},
  number={2},
  pages={1--18},
  year={2025},
  publisher={Springer}
}

@Article{Choi2020,
AUTHOR = {Choi, Younyoung and McClenen, Cayce},
TITLE = {Development of Adaptive Formative Assessment System Using Computerized Adaptive Testing and Dynamic Bayesian Networks},
JOURNAL = {Applied Sciences},
VOLUME = {10},
YEAR = {2020},
NUMBER = {22},
ARTICLE-NUMBER = {8196},
URL = {https://www.mdpi.com/2076-3417/10/22/8196},
ISSN = {2076-3417},
DOI = {10.3390/app10228196}
}

@article{yang2022,
title = {Adaptive formative assessment system based on computerized adaptive testing and the learning memory cycle for personalized learning},
journal = {Computers and Education: Artificial Intelligence},
volume = {3},
pages = {100104},
year = {2022},
issn = {2666-920X},
doi = {https://doi.org/10.1016/j.caeai.2022.100104},
url = {https://www.sciencedirect.com/science/article/pii/S2666920X22000595},
author = {Albert C.M. Yang and Brendan Flanagan and Hiroaki Ogata}
}

@article{wiggins2012seven,
  title={Seven keys to effective feedback},
  author={Wiggins, Grant},
  journal={Feedback},
  volume={70},
  number={1},
  pages={10--16},
  year={2012}
}

@article{hattie2007,
author = {John Hattie and Helen Timperley},
title ={The Power of Feedback},
journal = {Review of Educational Research},
volume = {77},
number = {1},
pages = {81-112},
year = {2007},
doi = {10.3102/003465430298487},
URL = {https://doi.org/10.3102/003465430298487 },
eprint = {   https://doi.org/10.3102/003465430298487}
}

@inproceedings{han2024,
author = {Han, Ariel and Zhou, Xiaofei and Cai, Zhenyao and Han, Shenshen and Ko, Richard and Corrigan, Seth and Peppler, Kylie A},
title = {Teachers, Parents, and Students' perspectives on Integrating Generative AI into Elementary Literacy Education},
year = {2024},
isbn = {9798400703300},
publisher = {Association for Computing Machinery},
address = {New York, NY, USA},
url = {https://doi.org/10.1145/3613904.3642438},
doi = {10.1145/3613904.3642438},
booktitle = {Proceedings of the 2024 CHI Conference on Human Factors in Computing Systems},
articleno = {678},
numpages = {17},
location = {Honolulu, HI, USA},
series = {CHI '24}
}

@inproceedings{prasad2025,
author = {Prasad, Prajish and Balse, Rishabh and Balchandani, Dhwani},
title = {Exploring Multimodal Generative AI for Education through Co-design Workshops with Students},
year = {2025},
isbn = {9798400713941},
publisher = {Association for Computing Machinery},
address = {New York, NY, USA},
url = {https://doi.org/10.1145/3706598.3714146},
doi = {10.1145/3706598.3714146},
booktitle = {Proceedings of the 2025 CHI Conference on Human Factors in Computing Systems},
articleno = {139},
numpages = {17},
location = {
},
series = {CHI '25}
}

@article{levin1980children,
  title={Children’s learning of all the news that’s fit to picture},
  author={Levin, Joel R and Berry, Jill K},
  journal={ECTJ},
  volume={28},
  number={3},
  pages={177--185},
  year={1980},
  publisher={Springer}
}

@article{Deci2000,
author = {Edward L. Deci and Richard M. Ryan},
title = {The "What" and "Why" of Goal Pursuits: Human Needs and the Self-Determination of Behavior},
journal = {Psychological Inquiry},
volume = {11},
number = {4},
pages = {227--268},
year = {2000},
publisher = {Routledge},
doi = {10.1207/S15327965PLI1104\_01},
URL = {https://doi.org/10.1207/S15327965PLI1104\_01
},
eprint = {   
        https://doi.org/10.1207/S15327965PLI1104\_01
}

}

@article{ryan2000self,
  title={Self-determination theory and the facilitation of intrinsic motivation, social development, and well-being.},
  author={Ryan, Richard M and Deci, Edward L},
  journal={American psychologist},
  volume={55},
  number={1},
  pages={68},
  year={2000},
  publisher={American Psychological Association}
}

@article{Zimmerman2009,
author = {Zimmerman, B.J. and Moylan, Adam},
year = {2009},
month = {01},
pages = {299-315},
title = {Self-regulation: Where metacognition and motivation intersect},
journal = {Handbook of metacognition in education}
}

@article{zimmerman2000,
title = {Self-Efficacy: An Essential Motive to Learn},
journal = {Contemporary Educational Psychology},
volume = {25},
number = {1},
pages = {82-91},
year = {2000},
issn = {0361-476X},
doi = {https://doi.org/10.1006/ceps.1999.1016},
url = {https://www.sciencedirect.com/science/article/pii/S0361476X99910160},
author = {Barry J. Zimmerman}
}

@ARTICLE{Panadero2017,
  
AUTHOR={Panadero, Ernesto },
         
TITLE={A Review of Self-regulated Learning: Six Models and Four Directions for Research},
        
JOURNAL={Frontiers in Psychology},
        
VOLUME={Volume 8 - 2017},

YEAR={2017},

URL={https://www.frontiersin.org/journals/psychology/articles/10.3389/fpsyg.2017.00422},

DOI={10.3389/fpsyg.2017.00422},

ISSN={1664-1078}}

@inproceedings{do2025,
author = {Do, Tiffany D. and Shafqat, Usama Bin and Ling, Elsie and Sarda, Nikhil},
title = {PAIGE: Examining Learning Outcomes and Experiences with Personalized AI-Generated Educational Podcasts},
year = {2025},
isbn = {9798400713941},
publisher = {Association for Computing Machinery},
address = {New York, NY, USA},
url = {https://doi.org/10.1145/3706598.3713460},
doi = {10.1145/3706598.3713460},
booktitle = {Proceedings of the 2025 CHI Conference on Human Factors in Computing Systems},
articleno = {896},
numpages = {12},
location = {
},
series = {CHI '25}
}

@techreport{unicef2019ai,
  title = {AI and Children: A Workshop in New York, 2019},
  author = {{UNICEF Innocenti}},
  institution = {UNICEF Innocenti},
  year = {2019},
  url = {https://www.unicef.org/innocenti/media/2486/file/AI-Children-Workshop-New-York-2019.pdf}
}

@article{vanmechelen2023,
author = {Van Mechelen, Maarten and Smith, Rachel Charlotte and Schaper, Marie-Monique and Tamashiro, Mariana and Bilstrup, Karl-Emil and Lunding, Mille and Graves Petersen, Marianne and Sejer Iversen, Ole},
title = {Emerging Technologies in K–12 Education: A Future HCI Research Agenda},
year = {2023},
issue_date = {June 2023},
publisher = {Association for Computing Machinery},
address = {New York, NY, USA},
volume = {30},
number = {3},
issn = {1073-0516},
url = {https://doi.org/10.1145/3569897},
doi = {10.1145/3569897},
journal = {ACM Trans. Comput.-Hum. Interact.},
month = jun,
articleno = {47},
numpages = {40}
}

@incollection{lumen_brain_adolescence,
  author = {Lumen Learning},
  title = {Brain Development During Adolescence},
  booktitle = {Lifespan Development},
  publisher = {Lumen Learning},
  year = {2019},
  url ={https://socialsci.libretexts.org/Bookshelves/Human_Development/Lifespan_Development_(Lumen)/07%3A_Adolescence/7.04%3A_Brain_Development_During_Adolescence.pdf},
    note = {This work is licensed under the Creative Commons Attribution 4.0 International License. http://creativecommons.org/licenses/by/4.0/}
}

@article{Pallant2025,
author = {Jessica L. Pallant and Janneke Blijlevens and Alexander Campbell and Ryan Jopp},
title = {Mastering knowledge: the impact of generative AI on student learning outcomes},
journal = {Studies in Higher Education},
volume = {0},
number = {0},
pages = {1--22},
year = {2025},
publisher = {SRHE Website},
doi = {10.1080/03075079.2025.2487570},
URL = {    
        https://doi.org/10.1080/03075079.2025.2487570
},
eprint = {    
        https://doi.org/10.1080/03075079.2025.2487570
}
}

@article{LUCZAK2024,
title = {Enhancing Academic Tutoring with AI – A Conceptual Framework},
journal = {Procedia Computer Science},
volume = {246},
pages = {5555-5564},
year = {2024},
note = {28th International Conference on Knowledge Based and Intelligent information and Engineering Systems (KES 2024)},
issn = {1877-0509},
doi = {https://doi.org/10.1016/j.procs.2024.09.709},
url = {https://www.sciencedirect.com/science/article/pii/S1877050924027704},
author = {Kamila Łuczak and Andrzej Greńczuk and Iwona Chomiak-Orsa and Estera Piwoni-Krzeszowska}
}

@ARTICLE{Zhai2021,
  title     = "A review of artificial intelligence ({AI}) in education from
               2010 to 2020",
  author    = "Zhai, Xuesong and Chu, Xiaoyan and Chai, Ching Sing and Jong,
               Morris Siu Yung and Istenic, Andreja and Spector, Michael and
               Liu, Jia-Bao and Yuan, Jing and Li, Yan",
  journal   = "Complexity",
  publisher = "Wiley",
  volume    =  2021,
  number    =  1,
  pages     = "1--18",
  month     =  jan,
  year      =  2021,
  copyright = "http://creativecommons.org/licenses/by/4.0/"
}

@article{ZHANG2021,
title = {AI technologies for education: Recent research \& future directions},
journal = {Computers and Education: Artificial Intelligence},
volume = {2},
pages = {100025},
year = {2021},
issn = {2666-920X},
doi = {https://doi.org/10.1016/j.caeai.2021.100025},
url = {https://www.sciencedirect.com/science/article/pii/S2666920X21000199},
author = {Ke Zhang and Ayse Begum Aslan}
}

@article{Adeshola2024,
author = {Ibrahim Adeshola and Adeola Praise Adepoju},
title = {The opportunities and challenges of ChatGPT in education},
journal = {Interactive Learning Environments},
volume = {32},
number = {10},
pages = {6159--6172},
year = {2024},
publisher = {Routledge},
doi = {10.1080/10494820.2023.2253858},


URL = {    
        https://doi.org/10.1080/10494820.2023.2253858
},
eprint = {     
        https://doi.org/10.1080/10494820.2023.2253858
}
}

@article{MEMARIAN2023,
title = {ChatGPT in education: Methods, potentials, and limitations},
journal = {Computers in Human Behavior: Artificial Humans},
volume = {1},
number = {2},
pages = {100022},
year = {2023},
issn = {2949-8821},
doi = {https://doi.org/10.1016/j.chbah.2023.100022},
url = {https://www.sciencedirect.com/science/article/pii/S2949882123000221},
author = {Bahar Memarian and Tenzin Doleck}
}

@inproceedings{Li2025,
author = {Li, Tiffany Wenting and Song, Yifan and Sundaram, Hari and Karahalios, Karrie},
title = {Can Learners Navigate Imperfect Generative Pedagogical Chatbots? An Analysis of Chatbot Errors on Learning},
year = {2025},
isbn = {9798400712913},
publisher = {Association for Computing Machinery},
address = {New York, NY, USA},
url = {https://doi.org/10.1145/3698205.3729550},
doi = {10.1145/3698205.3729550},
booktitle = {Proceedings of the Twelfth ACM Conference on Learning @ Scale},
pages = {151–163},
numpages = {13},
location = {Palermo, Italy},
series = {L@S '25}
}

@ARTICLE{Moreno2007,
  title     = "Interactive multimodal learning environments",
  author    = "Moreno, Roxana and Mayer, Richard",
  journal   = "Educ. Psychol. Rev.",
  publisher = "Springer Science and Business Media LLC",
  volume    =  19,
  number    =  3,
  pages     = "309--326",
  month     =  sep,
  year      =  2007
}

@misc{dennison2025,
      title={Teacher-AI Collaboration for Curating and Customizing Lesson Plans in Low-Resource Schools}, 
      author={Deepak Varuvel Dennison and Bakhtawar Ahtisham and Kavyansh Chourasia and Nirmit Arora and Rahul Singh and Rene F. Kizilcec and Akshay Nambi and Tanuja Ganu and Aditya Vashistha},
      year={2025},
      eprint={2507.00456},
      archivePrefix={arXiv},
      primaryClass={cs.CY},
      url={https://arxiv.org/abs/2507.00456}, 
}

@inproceedings{Park2024,
author = {Park, Hyanghee and Ahn, Daehwan},
title = {The Promise and Peril of ChatGPT in Higher Education: Opportunities, Challenges, and Design Implications},
year = {2024},
isbn = {9798400703300},
publisher = {Association for Computing Machinery},
address = {New York, NY, USA},
url = {https://doi.org/10.1145/3613904.3642785},
doi = {10.1145/3613904.3642785},
booktitle = {Proceedings of the 2024 CHI Conference on Human Factors in Computing Systems},
articleno = {271},
numpages = {21},
location = {Honolulu, HI, USA},
series = {CHI '24}
}

@inproceedings{Tankelevitch2025,
author = {Tankelevitch, Lev and Glassman, Elena L. and He, Jessica and Kazemitabaar, Majeed and Kittur, Aniket and Lee, Mina and Palani, Srishti and Sarkar, Advait and Ramos, Gonzalo and Rogers, Yvonne and Subramonyam, Hari},
title = {Tools for Thought: Research and Design for Understanding, Protecting, and Augmenting Human Cognition with Generative AI},
year = {2025},
isbn = {9798400713958},
publisher = {Association for Computing Machinery},
address = {New York, NY, USA},
url = {https://doi.org/10.1145/3706599.3706745},
doi = {10.1145/3706599.3706745},
booktitle = {Proceedings of the Extended Abstracts of the CHI Conference on Human Factors in Computing Systems},
articleno = {804},
numpages = {8},
location = {
},
series = {CHI EA '25}
}

@ARTICLE{Chan2023,
  title     = "The {AI} generation gap: Are Gen {Z} students more interested in
               adopting generative {AI} such as {ChatGPT} in teaching and
               learning than their Gen {X} and millennial generation teachers?",
  author    = "Chan, Cecilia Ka Yuk and Lee, Katherine K W",
  journal   = "Smart Learn. Environ.",
  publisher = "Springer Science and Business Media LLC",
  volume    =  10,
  number    =  1,
  month     =  nov,
  year      =  2023,
  copyright = "https://creativecommons.org/licenses/by/4.0"
}

@misc{denny2023,
      title={Can We Trust AI-Generated Educational Content? Comparative Analysis of Human and AI-Generated Learning Resources}, 
      author={Paul Denny and Hassan Khosravi and Arto Hellas and Juho Leinonen and Sami Sarsa},
      year={2023},
      eprint={2306.10509},
      archivePrefix={arXiv},
      primaryClass={cs.HC},
      url={https://arxiv.org/abs/2306.10509}, 
}

@ARTICLE{Mayer1990,
  title     = "When is an illustration worth ten thousand words?",
  author    = "Mayer, Richard E and Gallini, Joan K",
  journal   = "J. Educ. Psychol.",
  publisher = "American Psychological Association (APA)",
  volume    =  82,
  number    =  4,
  pages     = "715--726",
  month     =  dec,
  year      =  1990
}

@ARTICLE{Abdaljaleel2024,
  title    = "A multinational study on the factors influencing university
              students' attitudes and usage of {ChatGPT}",
  author   = "Abdaljaleel, Maram and Barakat, Muna and Alsanafi, Mariam and
              Salim, Nesreen A and Abazid, Husam and Malaeb, Diana and
              Mohammed, Ali Haider and Hassan, Bassam Abdul Rasool and Wayyes,
              Abdulrasool M and Farhan, Sinan Subhi and Khatib, Sami El and
              Rahal, Mohamad and Sahban, Ali and Abdelaziz, Doaa H and Mansour,
              Noha O and AlZayer, Reem and Khalil, Roaa and Fekih-Romdhane,
              Feten and Hallit, Rabih and Hallit, Souheil and Sallam, Malik",
  journal  = "Sci. Rep.",
  volume   =  14,
  number   =  1,
  pages    = "1983",
  month    =  jan,
  year     =  2024
}

@inproceedings{Do2022,
author = {Do, Tiffany D. and Akter, Mamtaj and Choudhary, Zubin and Azevedo, Roger and McMahan, Ryan P.},
title = {The Effects of an Embodied Pedagogical Agent’s Synthetic Speech Accent on Learning Outcomes},
year = {2022},
isbn = {9781450393904},
publisher = {Association for Computing Machinery},
address = {New York, NY, USA},
url = {https://doi.org/10.1145/3536221.3556587},
doi = {10.1145/3536221.3556587},
booktitle = {Proceedings of the 2022 International Conference on Multimodal Interaction},
pages = {198–206},
numpages = {9},
location = {Bengaluru, India},
series = {ICMI '22}
}

@book{bloom1971taxonomy,
  title={Taxonomy of educational objectives: The classification of educational goals: By a committee of college and university examiners},
  author={Bloom, Benjamin S},
  year={1971},
  publisher={David McKay}
}

@article{patall2008effects,
  title={The effects of choice on intrinsic motivation and related outcomes: a meta-analysis of research findings.},
  author={Patall, Erika A and Cooper, Harris and Robinson, Jorgianne Civey},
  journal={Psychological bulletin},
  volume={134},
  number={2},
  pages={270},
  year={2008},
  publisher={American Psychological Association}
}

@techreport{LYWtechreport,
  author    = "Amy Wang and Anna Iurchenko and Anisha Choudhury and Alicia Martín and Amir Globerson and Avinatan Hassidim and Ayça Çakmakli and Ayelet Shasha Evron and Charlie Yang and Courtney Heldreth and Diana Akrong and Gal Elidan and Hairong Mu and Ian Li and Ido Cohen and Katherine Chou and Komal Singh and Lev Borovoi and Lidan Hackmon and Lior Belinsky and Michael Fink and Niv Efron and Preeti Singh and Rena Levitt and Shashank Agarwal and Shay Sharon and Tracey Lee-Joe and Xiaohong Hao and Yael Gold-Zamir and Yael Haramaty and Yishay Mor and Yoav Bar Sinai and Yossi Matias",
  title     = "Towards an AI-Augmented Textbook",
  institution = "Google Research",
  year      = "2025",
      eprint={2509.13348},
      archivePrefix={arXiv},
      primaryClass={cs.CY},
      url={https://arxiv.org/abs/2509.13348}, 
}

@article{Hothorn2008,
 title={Implementing a Class of Permutation Tests: The coin Package},
 volume={28},
 url={https://www.jstatsoft.org/index.php/jss/article/view/v028i08},
 doi={10.18637/jss.v028.i08},
 number={8},
 journal={Journal of Statistical Software},
 author={Hothorn, Torsten and Hornik, Kurt and van de Wiel, Mark A. and Zeileis, Achim},
 year={2008},
 pages={1–23}
}

@ARTICLE{Shapiro1965,
  title     = "An analysis of variance test for normality (complete samples)",
  author    = "Shapiro, S S and Wilk, M B",
  journal   = "Biometrika",
  publisher = "Oxford University Press (OUP)",
  volume    =  52,
  number    = "3-4",
  pages     = "591--611",
  month     =  dec,
  year      =  1965
}

@book{braun2021thematic,
  title={Thematic analysis: A practical guide},
  author={Braun, Virginia and Clarke, Victoria},
  year={2021},
  publisher={SAGE publications Ltd}
}

@article{clark1991dual,
  title={Dual coding theory and education},
  author={Clark, James M and Paivio, Allan},
  journal={Educational psychology review},
  volume={3},
  number={3},
  pages={149--210},
  year={1991},
  publisher={Springer}
}

@book{Paivio1990,
    author = {Paivio, Allan},
    title = {Mental Representations: A dual coding approach},
    publisher = {Oxford University Press},
    year = {1990},
    month = {09},
    isbn = {9780195066661},
    doi = {10.1093/acprof:oso/9780195066661.001.0001},
    url = {https://doi.org/10.1093/acprof:oso/9780195066661.001.0001},
}

@INPROCEEDINGS{ChangChien2024,
  author={Chang, Chia-Kai and Chien, Lee-Chia-Tung},
  booktitle={2024 IEEE International Conference on Advanced Learning Technologies (ICALT)}, 
  title={Enhancing Academic Performance with Generative AI-Based Quiz Platform}, 
  year={2024},
  volume={},
  number={},
  pages={193-195},
  doi={10.1109/ICALT61570.2024.00062}
}

@article{al2022perceptions,
  title={Perceptions of learners and instructors towards artificial intelligence in personalized learning},
  author={Al-Badi, Ali and Khan, Asharul and others},
  journal={Procedia computer science},
  volume={201},
  pages={445--451},
  year={2022},
  publisher={Elsevier}
}

@article{Kulik2016,
author = {James A. Kulik and J. D. Fletcher},
title ={Effectiveness of Intelligent Tutoring Systems: A Meta-Analytic Review},

journal = {Review of Educational Research},
volume = {86},
number = {1},
pages = {42-78},
year = {2016},
doi = {10.3102/0034654315581420},

URL = { 
    
        https://doi.org/10.3102/0034654315581420
    
    

},
eprint = { 
    
        https://doi.org/10.3102/0034654315581420
    
    

}
}

@article{SCHNEIDER2018,
title = {A meta-analysis of how signaling affects learning with media},
journal = {Educational Research Review},
volume = {23},
pages = {1-24},
year = {2018},
issn = {1747-938X},
doi = {https://doi.org/10.1016/j.edurev.2017.11.001},
url = {https://www.sciencedirect.com/science/article/pii/S1747938X17300581},
author = {Sascha Schneider and Maik Beege and Steve Nebel and Günter Daniel Rey}
}

@article{Moreno2000,
author = {Moreno, Roxana and Mayer, Richard},
year = {2000},
month = {12},
pages = {724-733},
title = {Engaging Students in Active Learning: The Case for Personalized Multimedia Messages},
volume = {92},
journal = {Journal of Educational Psychology},
doi = {10.1037/0022-0663.92.4.724}
}

@ARTICLE{Coens2011,
  title     = "Listening to an educational podcast while walking or jogging",
  author    = "Coens, Joke and Degryse, Ellen and Senecaut, Marie-Paul and
               Cottyn, Jorge and Clarebout, Geraldine",
  journal   = "Int. J. Mob. Blended Learn.",
  publisher = "IGI Global",
  volume    =  3,
  number    =  3,
  pages     = "23--33",
  month     =  jul,
  year      =  2011
}

@ARTICLE{Sun2024,
  title     = "Would {ChatGPT-facilitated} programming mode impact college
               students' programming behaviors, performances, and perceptions?
               An empirical study",
  author    = "Sun, Dan and Boudouaia, Azzeddine and Zhu, Chengcong and Li, Yan",
  journal   = "Int. J. Educ. Technol. High. Educ.",
  publisher = "Springer Science and Business Media LLC",
  volume    =  21,
  number    =  1,
  month     =  feb,
  year      =  2024,
  copyright = "https://creativecommons.org/licenses/by/4.0"
}

@misc{mccall2024future,
  title={The Future of Learning Styles: Emerging Trends and Technologies},
  author={MCcall, Andrei},
  year={2024},
  publisher={ResearchGate}
}

@ARTICLE{Villegas2024,
  author={Villegas-Ch, William and García-Ortiz, Joselin and Sánchez-Viteri, Santiago},
  journal={IEEE Access}, 
  title={Personalization of Learning: Machine Learning Models for Adapting Educational Content to Individual Learning Styles}, 
  year={2024},
  volume={12},
  number={},
  pages={121114-121130},
  doi={10.1109/ACCESS.2024.3452592}}

@article{frank2024influence,
  title={The Influence of artificial intelligence on education: Enhancing personalized learning experiences},
  author={Frank, Edwin},
  journal={EasyChair Preprint},
  volume={14675},
  year={2024}
}

@ARTICLE{Cordova1996,
  title     = "Intrinsic motivation and the process of learning: Beneficial
               effects of contextualization, personalization, and choice",
  author    = "Cordova, Diana I and Lepper, Mark R",
  journal   = "J. Educ. Psychol.",
  publisher = "American Psychological Association (APA)",
  volume    =  88,
  number    =  4,
  pages     = "715--730",
  month     =  dec,
  year      =  1996
}

@article{schneider2018autonomy,
  title={The autonomy-enhancing effects of choice on cognitive load, motivation and learning with digital media},
  author={Schneider, Sascha and Nebel, Steve and Beege, Maik and Rey, G{\"u}nter Daniel},
  journal={Learning and Instruction},
  volume={58},
  pages={161--172},
  year={2018},
  publisher={Elsevier}
}

@article{surjono2015effects,
  title={The effects of multimedia and learning style on student achievement in online electronics course.},
  author={Surjono, Herman Dwi},
  journal={Turkish Online Journal of Educational Technology-TOJET},
  volume={14},
  number={1},
  pages={116--122},
  year={2015},
  publisher={ERIC}
}

@ARTICLE{Popescu2010,
  title     = "Adaptation provisioning with respect to learning styles in a
               Web‐based educational system: an experimental study",
  author    = "Popescu, E",
  journal   = "J. Comput. Assist. Learn.",
  publisher = "Wiley",
  volume    =  26,
  number    =  4,
  pages     = "243--257",
  month     =  aug,
  year      =  2010
}

@article{hennink2017code,
  title={Code saturation versus meaning saturation: how many interviews are enough?},
  author={Hennink, Monique M and Kaiser, Bonnie N and Marconi, Vincent C},
  journal={Qualitative health research},
  volume={27},
  number={4},
  pages={591--608},
  year={2017},
  publisher={Sage Publications Sage CA: Los Angeles, CA}
}

@article{shulman1987knowledge,
  title={Knowledge and teaching: Foundations of the new reform},
  author={Shulman, Lee},
  journal={Harvard educational review},
  volume={57},
  number={1},
  pages={1--23},
  year={1987},
  publisher={Harvard Education Publishing Group}
}

@inproceedings{Koedinger2015,
author = {Koedinger, Kenneth R. and Kim, Jihee and Jia, Julianna Zhuxin and McLaughlin, Elizabeth A. and Bier, Norman L.},
title = {Learning is Not a Spectator Sport: Doing is Better than Watching for Learning from a MOOC},
year = {2015},
isbn = {9781450334112},
publisher = {Association for Computing Machinery},
address = {New York, NY, USA},
url = {https://doi.org/10.1145/2724660.2724681},
doi = {10.1145/2724660.2724681},
booktitle = {Proceedings of the Second (2015) ACM Conference on Learning @ Scale},
pages = {111–120},
numpages = {10},
location = {Vancouver, BC, Canada},
series = {L@S '15}
}

@article{brusilovsky2001adaptive,
  title={Adaptive hypermedia},
  author={Brusilovsky, Peter},
  journal={User modeling and user-adapted interaction},
  volume={11},
  number={1},
  pages={87--110},
  year={2001},
  publisher={Springer}
}

@article{mitrovic2012fifteen,
  title={Fifteen years of constraint-based tutors: what we have achieved and where we are going},
  author={Mitrovic, Antonija},
  journal={User modeling and user-adapted interaction},
  volume={22},
  number={1},
  pages={39--72},
  year={2012},
  publisher={Springer}
}

@book{ebbinghaus1913memory,
  title={Memory: A Contribution to Experimental Psychology},
  author={Ebbinghaus, Hermann},
  translator={Ruger, Henry A. and Bussenius, Clara E.},
  year={1913},
  publisher={Teachers College, Columbia University},
  address={New York},
  note={Original work published 1885}
}

@book{wood2017gam,
  title={Generalized Additive Models: An Introduction with R},
  author={Wood, Simon N.},
  year={2017},
  edition={2nd},
  publisher={CRC Press},
  address={Boca Raton, FL}
}

@article{benjamini1995controlling,
  title={Controlling the false discovery rate: a practical and powerful approach to multiple testing},
  author={Benjamini, Yoav and Hochberg, Yosef},
  journal={Journal of the Royal Statistical Society: Series B (Methodological)},
  volume={57},
  number={1},
  pages={289--300},
  year={1995},
  publisher={Wiley Online Library}
}


\appendix
\section*{Appendix}
\section{Personalization} \label{appA}
\begin{figure*}[h]
\centering
  \includegraphics[width=0.6\textwidth]{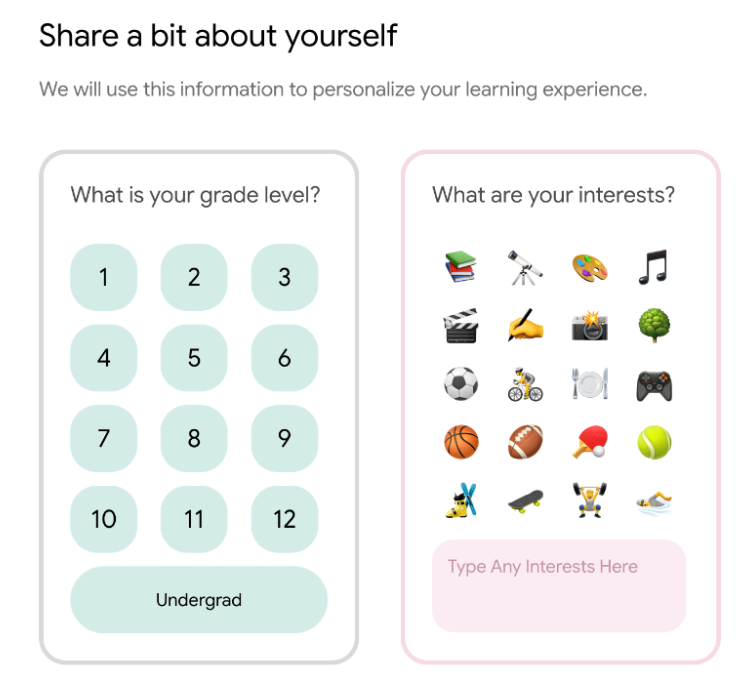}
  \caption{Visual of how students are able to personalize based on grade level and student interest in \textit{Learn Your Way}}
  \label{fig:figurePersonalization}
\end{figure*}

\end{document}